\documentclass[useAMS,usenatbib]{mn2e}

\pdfoutput=1

\usepackage{amssymb}
\usepackage{amsmath}
\usepackage{graphicx}
\usepackage{epstopdf}
\usepackage{subfigure}
\usepackage{aasmacros} 
\usepackage{multirow}
\usepackage{pbox}
\usepackage{url}

\usepackage{xspace}

\def\gtsim{>\kern-1.2em\lower1.1ex\hbox{$\sim$}~}   
\def\ltsim{<\kern-1.2em\lower1.1ex\hbox{$\sim$}~}   
\def\ltsima{$\; \buildrel < \over \sim \;$}
\def\simlt{\lower.5ex\hbox{\ltsima}}
\def\gtsima{$\; \buildrel > \over \sim \;$}
\def\simgt{\lower.5ex\hbox{\gtsima}}

\newcommand{\fig}[1]{Fig.~\ref{#1}}
\newcommand{\tab}[1]{Table~\ref{#1}}
\newcommand{\sect}[1]{Section~\ref{#1}}

\def\subfind{\textsc{subfind}\xspace}

\newcommand{\gadgetthree}[0]{{\sc gadget-3}\xspace}

\def\etal{{\it et al.~\/}}

\def\lcdm{$\Lambda$CDM\xspace}

\def\kms { {kms$^{-1}$} \xspace}

\def\kpc{$h^{-1}$~kpc\xspace}

\def\kms{{\rm\,km\,s^{-1}}}

\def\hide#1{}

\usepackage{xparse}

\def\msol{\mathrm{M_{\sun}}}
\def\kpc{\,\mathrm{kpc}}
\def\rtwohundred{r_{200}}
\def\acc{{accreted}\xspace}    
\def\eris{\textit{Eris}}    
\def\ins{{\it in situ}\xspace} 
\def\aqc{{Aq-C}\xspace}
\def\aqd{{Aq-D}\xspace}
\def\aqe{{Aq-E}\xspace}
\def\kms{\,\mathrm{km \, s^{-1}}}

\def\tiss{Tissera et al.}
\def\ttwelve{T12}
\def\tthirteen{T13}

\newcommand{\REF}[1]{#1}
\newcommand{\REFB}[1]{#1}

\title[In Situ Stellar Haloes]{Formation of In Situ Stellar Haloes in Milky Way-Mass Galaxies}
\author[Cooper \etal]{Andrew P. Cooper$^{1}$\thanks{E-mail:
a.p.cooper@durham.ac.uk}, Owen H. Parry$^{2}$, Ben Lowing$^{1}$, Shaun Cole$^{1}$ and Carlos
Frenk$^{1}$ \\
$^{1}$Institute for Computational Cosmology, Department of Physics, University of Durham, South Road, Durham, DH1 3LE, UK \\
$^{2}$Department of Astronomy, University of Maryland, College Park, MD 20742, USA}

\begin{document}

\date{Accepted 2015 September 03. Received 2015 August 11; in original form 2015 January 21}
\pagerange{\pageref{firstpage}--\pageref{lastpage}} \pubyear{2013}
\maketitle


\label{firstpage}

\begin{abstract} We study the formation of stellar haloes in three Milky
  Way-mass galaxies using cosmological smoothed particle hydrodynamics simulations, focusing on the subset
  of halo stars that form \textit{in situ}, as opposed to those accreted from
  satellites. \textit{In situ} stars in our simulations dominate the stellar halo out to
  20~kpc and account for 30-40 per cent of its total mass. We separate \textit{in
  situ} halo stars into three straightforward, physically distinct categories
  according to their origin: stars scattered from the disc of the main galaxy
  (`heated disc'), stars formed from gas smoothly accreted on to the halo
  (`smooth' gas) and stars formed in streams of gas stripped from infalling
  satellites (`stripped' gas). We find that most belong to the stripped gas
  category. Those originating in smooth gas outside the disc tend to form at
  the same time and place as the stripped-gas population, suggesting that their
  formation is associated with the same gas-rich accretion events.  The
  scattered disc star contribution is negligible overall but significant in the
  Solar neighbourhood, where $\gtrsim 90$ per cent of stars on eccentric orbits
  once belonged to the disc. However, the distinction between halo and thick
  disc in this region is highly ambiguous.  The chemical and kinematic
  properties of the different components are very similar at the present day,
  but the global properties of the \textit{in situ} halo differ substantially between
  the three galaxies in our study. In our simulations, the
  hierarchical buildup of structure is the driving force behind not only the
  accreted stellar halo, but also those halo stars formed \textit{in situ}.

\end{abstract}

\begin{keywords}
  methods: numerical -- galaxies:formation -- galaxies: haloes-- galaxies: structure
\end{keywords}

\section{Introduction}  \label{sec:INTRO}

Following \citet{Searle78}, much observational and theoretical work on the Milky
Way's stellar halo has focused on the tidal stripping and disruption of satellite galaxies. The idea that galactic stellar haloes are built mainly by accretion is well supported by theoretical predictions of the standard dark energy/cold DM ($\Lambda$CDM) cosmogony \citep{WhiteFrenk91,Bullock05,Cooper10} and direct evidence of tidal streams around nearby galaxies \citep{Belokurov06, McConnachie09,
MartinezDelgado10}.  However, recent work has shown that some aspects of the
Milky Way's stellar halo may be difficult to explain by accretion alone,
notably its central concentration and uniformity across the sky \citep{Carollo2007, Bell08,
Cooper11a,, Deason2011, Helmi11,  Xue11}.

Meanwhile, hydrodynamical simulations have predicted a distinct `\ins' halo
component, defined (loosely) as having formed bound to the Milky Way itself
rather than to any of its hierarchical progenitors \citep{Abadi2003, Brook04b,
Zolotov2009, Font2011b, Tissera13a, Pillepich15a}. Such haloes are a natural
outcome of the $\Lambda$CDM model, which predicts that the vast majority of
stars in a galaxy like the Milky Way form from the cooling of gas trapped by
the galaxy's own DM (DM) potential \citep{WhiteRees1978,WhiteFrenk91}. The
bulk of these \ins stars can be identified with the kinematically cold, rapidly
rotating Galactic disc, but the proto-Milky Way may also have suffered strong
perturbations from satellites and quasi-secular rearrangement \citep[e.g.
`disc flips';][]{Bett12}, or even wholesale destruction and regrowth before the
majority of present-day disc stars were formed \citep[e.g.][]{Sales12,
Aumer13b, Aumer13a}. If real galaxies pass through such messy stages of
formation, it seems likely that a significant fraction of stars formed in the
early Galaxy would now have highly eccentric orbits.  

Recent simulations find that these \ins processes create haloes that are more
concentrated, metal-rich and oblate than those formed by \acc stars
\citep{McCarthy12}.  This supports the hypothesis of a transition between an
\ins and an \acc halo as an explanation for the apparently `bimodal' properties
of halo stars observed in the Milky Way \citep{Carollo10, Beers12, Tissera14}.

However, quantitative results concerning the origin of \ins halo stars and
their importance relative to \acc stars are still very uncertain. Where they
rely on simulations, such conclusions can be particularly sensitive to the
numerical methods used. Starting from identical initial conditions, current
state-of-the-art simulations predict substantially different properties for the
bulk of the \ins stellar mass in Milky Way-like DM haloes
\citep[e.g.][]{Aquila, Aumer13a}, not just the few per cent that might be
identified with an \ins halo. Moreover, the properties of \ins haloes may be
much more sensitive to certain modelling choices than those of massive stellar
discs, including prescriptions for star formation and the treatment of the
multi-phase interstellar medium (ISM; for example, the mixing of hot and cold in
galactic winds, tidal streams and cold clumps of free-falling gas).

Here we analyse the origin of \ins halo stars using three Milky Way-scale
simulations run with the code described in \citet{Parry2012}, one of the
participants in the Aquila comparison project \citep{Aquila}.  Two of the three
DM haloes we simulate have also been simulated by \citet[][hearafter
T12]{Tissera12} and \citet[][hearafter T13, T14]{Tissera13a,Tissera14} using
different `subgrid' recipes for star formation and feedback but an otherwise
similar hydrodynamic solver and identical initial conditions. We define what we
mean by \ins halo stars in a straightforward and easily reproducible way.
Careful definitions are particularly important for this problem because the
concept of an \ins halo straddles an extremely fuzzy boundary between all the
conventional Galactic components -- disc, thick disc, bulge and halo.
\REF{Based on these definitions, we discuss physical mechanisms by which \ins
haloes are generated in our simulations, motivated by the fact that the
mechanisms we identify are somewhat different from those previously discussed
in the literature. In particular, we find that the growth of the \ins halo is
very closely related to the accretion of satellites responsible for the growth
of the canonical `\textit{ex situ}' halo.}

We proceed as follows. We describe our simulations in Section~\ref{sec:SIM}.
In Section~\ref{sec:SAMPLE} we explain how we identify \ins halo stars and in
Section~\ref{sec:ORIGIN} we examine their origins.  Section~\ref{sec:PROPTODAY}
describes the present-day properties of our \ins halo.
Section~\ref{sec:PROGENITORS} investigates the satellite progenitor of \ins
stars formed from stripped gas. In Section~\ref{sec:DISCUSSION} we interpret
our results, discuss the limitations of \ins halo models based on
hydrodynamical simulations and compare with similar studies. A summary of our
conclusions is given in Section~\ref{sec:CONCLUSIONS}. \REF{A detailed
comparison with the results of \ttwelve{} and \tthirteen{} is included in
Appendices A and B, along with a short discussion of numerical convergence.}

\section{Simulations} \label{sec:SIM}

We examine the stellar haloes that form in three smoothed particle hydrodynamics
(SPH) simulations of Milky Way-mass galaxies.  Dark-matter-only versions of
these simulations formed part of the Aquarius project \citep{Springel2008} and
we retain the Aquarius nomenclature for our three sets of initial conditions,
labelling them \aqc, \aqd and \aqe.  The DM resolution (particle mass)
in our SPH simulations is similar to that of the `level 4' simulation set in
Aquarius. 

The Aquarius DM haloes were themselves extracted from a cosmological
simulation in a cube of comoving volume $\rm 100^{3} \, Mpc^{3}$.  They were
chosen to have masses close to that of the Milky Way ($\sim10^{12} \msol$) and
to avoid dense environments \citep[no neighbour exceeding half the mass of the
target halo within $1h^{-1}$Mpc;][]{Navarro2010}. Initial conditions for a
resimulation of each halo were created with a `zoom' technique, with higher
mass boundary particles used to model the large scale potential and lower mass
particles in an $\sim5 h^{-1}$Mpc region surrounding the target halo.  Extra
power was added to the initial particle distribution on small scales in the
high resolution region, as described by \citet{Frenk1996}. The numerical
parameters for each simulation, including the particle masses and gravitational
softening lengths, are listed in \tab{tab:SIM_PROPS}. We assume a \lcdm
cosmology, with parameters $\Omega_{\rm m}=0.25$, $\Omega_{\Lambda}=0.75$,
$\Omega_{\rm b}=0.045$, $\sigma_{8}=0.9$, $n_{\rm s}=1$ and $H_{0}=100h \, {\rm
km \, s}^{-1} \, {\rm Mpc}^{-1} = 73 \, {\rm km \, s}^{-1} \, {\rm Mpc}^{-1}$. 

\begin{table}
  \centering
  \caption{Numerical parameters adopted for the three simulations: DM and gas particle masses and the maximum gravitational softening length in physical units (defined as the scale of a Plummer kernel equivalent to the actual spline kernel used in the simulation).}
  \label{tab:SIM_PROPS}
  
  \begin{tabular}{cccc}
    &$\rm M_{\mathrm{DM}}(\msol)$& $\rm M_{\mathrm{gas}}(\msol)$&$\rm
    \epsilon_{phys}[pc]$\\ \hline \hline
    Aq-C&$2.6\times10^5$&$5.8\times10^4$&257\\
    Aq-D&$2.2\times10^5$&$4.8\times10^4$&257\\
    Aq-E&$2.1\times10^5$&$4.7\times10^4$&257\\ \hline
  \end{tabular}

\end{table} 

Our simulation code is based on an early version of the PM-Tree-SPH code
\gadgetthree.  Baryon processes are modelled as described in
\citet{Okamoto2010a} and \citet{Parry2012}. \REF{Briefly, each gas particle
represents an ISM with separate `hot' and `cold'
(star-forming) phases. Gas particles above a critical density ($n_{\mathrm{H}} > 0.1
\mathrm{\,cm^{-3}}$) are assigned a cold phase mass according to their thermal
energy and a local pressure according to a polytropic equation of state. Gas in
the cold phase is converted to stars at a rate inversely proportional to a
local dynamical time, which in turn depends on analytic approximations for the
effective pressure and distribution of cold cloud sizes}. Our code follows the
nucleosynthetic production of individual elements separately, in particular the
iron yields of Types II and I SNe.  \REF{Mass, metals and energy returned by
evolved stars and their SNe are smoothly distributed over 40 near
neighbour gas particles.  The effective metallicity used to calculate the
radiative cooling rate of a given particle is also a smoothed average. Kinetic
energy is imparted directly to particles subject to SNe feedback, in
proportion to the local velocity dispersion of DM. Particles are
entrained in SN winds on a probabilistic basis and launched
perpendicular to the plane of the galactic disc. Wind particles are decoupled
from the hydrodynamical calculation until they reach an ambient density $n_{\mathrm{H}}
< 0.01 \mathrm{\,cm^{-3}}$ or the time they have been decoupled exceeds a
limiting time.} \REF{\citet{Parry2012} improve the treatment of fluid
instabilities in the hydrodynamic scheme of \citet{Okamoto2010a} by
incorporating artificial conductivity \citep{Price2008} and a time-step limiter
\citep{Saitoh:2009aa}; they also adjust the treatment of SN winds to
reduce the resolution dependence of mass loading and deposit thermal rather
than kinetic energy from Type Ia SNe. The most relevant effect of
these improvements is to suppress the formation of stars in cooling
instabilities in diffuse circumgalactic gas \citep[e.g.][]{Kaufmann06,Keres12,
Hobbs13}. In the results we present below, we explicitly identify stars formed
through such instabilities and find that they contribute $\lesssim20$~per cent
of the mass of the \ins stellar halo in our simulations. \REFB{This fraction
depends somewhat on resolution (becoming larger at lower resolution) as we
discuss in Appendix B}. Other modelling uncertainties are also discussed
\REFB{in Section~\ref{sec:DISCUSSION} and Appendix B.}}

Details of two of our three hydrodynamical simulations (\aqc and \aqd) have
been presented previously in studies that focused on satellite galaxies
\citep{Parry2012} and pseudo-bulge formation \citep{Okamoto2013}.  The
luminosity function and luminosity--metallicity relation of the simulated
satellites are comparable to those of dwarf galaxies in the Local Group.  All
three simulations result in galaxies with massive centrifugally supported discs
as well as dispersion supported spheroids.  

\section{Sample definition} \label{sec:SAMPLE}

The first step in defining our stellar halo sample is to identify all stars
belonging to the central (Milky Way-analogue) galaxy at the present day
(redshift $z=0$). We choose stars that lie within a radius $\rtwohundred$ which
encloses a sphere of mean density 200 times the critical value for closure
($\rtwohundred=227$~kpc for \aqc and \aqd, 202~kpc for \aqe). From this sample,
we isolate the halo by excluding stars that belong to satellite galaxies within
$\rtwohundred$ and stars that belong to the main galaxy disc or \REF{an inner
spheroid (`bulge')}, as follows.


Satellite DM haloes and their galaxies are isolated using a version of the \subfind\ algorithm \citep{Springel2001} adapted by \citet{Dolag2009} to identify self-bound substructures, taking into account the internal energy of the gas when computing particle binding energies. All star particles bound to DM subhaloes at $z=0$ are excluded from our halo star sample.

The central galactic disc is identified by finding stars on orbits that are approximately circular and that lie close to a plane normal to the net angular momentum vector of the whole stellar component. A coordinate system is chosen such that the net angular momentum vector of all stars within $0.2 \rtwohundred$ points in the positive $z$-direction.  The circularity of each star's orbit is then defined as
\begin{equation}
\label{eqn:CIRC_DEF}
\mathcal{E}_{\rm E} = \frac{J_z}{J_{\rm circ}(E)},
\end{equation} where $J_z$ is the $z$ component of the star's specific angular momentum and $J_{\mathrm{circ}}(E)$ is the specific angular momentum of a star with the same binding energy on a circular orbit. All stars with $\mathcal{E}_{\rm E} > 0.8$ are identified with the central galactic disc and excluded from our halo star sample.

\begin{figure*}
  \centerline{\includegraphics[width=1.0\linewidth, clip=True, trim=2cm 0 0 0 ]{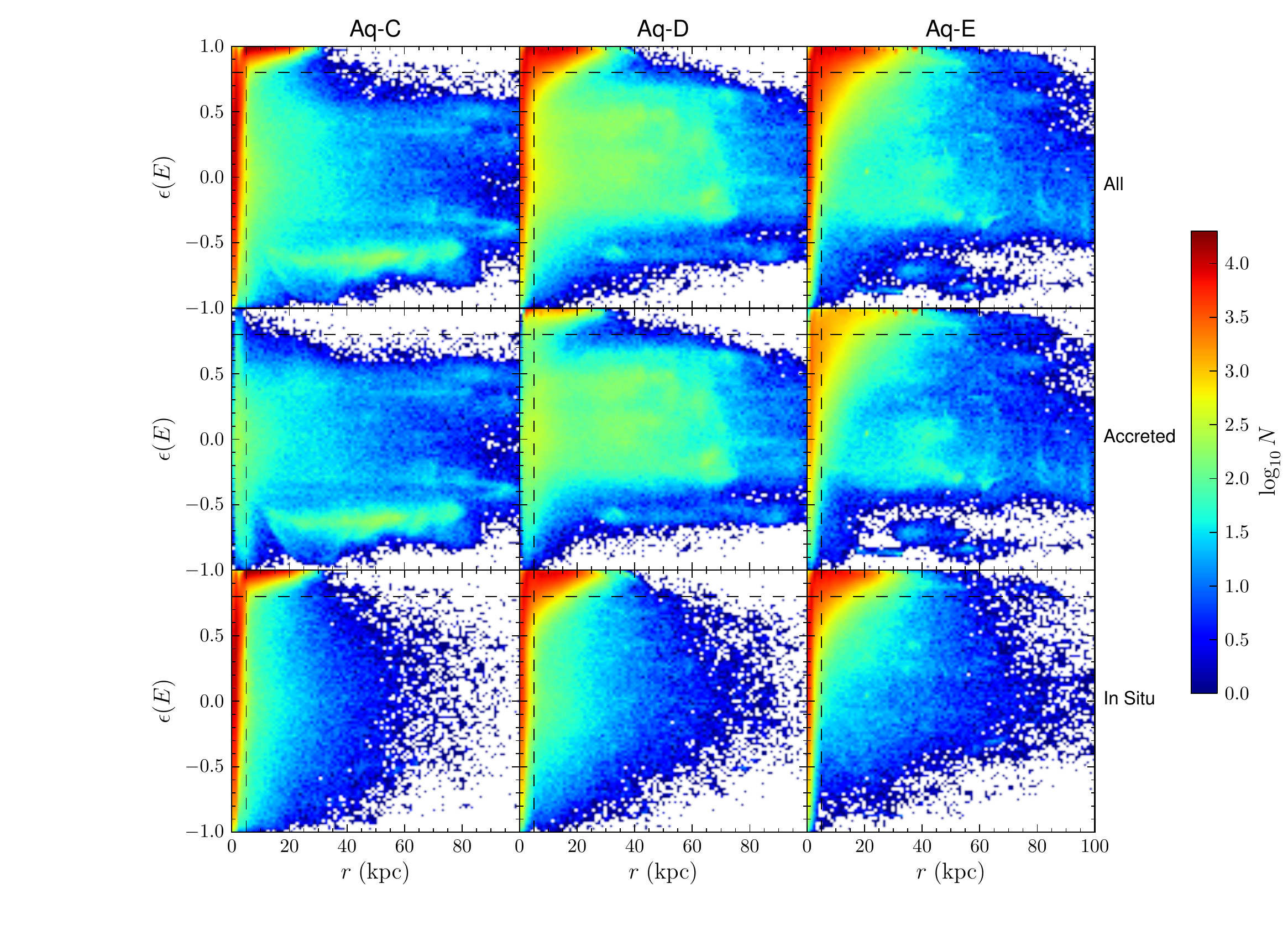}}

\caption{The distribution of stars in radius-circularity space for \aqc (left
column), \aqd (centre column) and \aqe (right column).  Panels in the top row
include all stars bound to the main DM halo ($r<90$~kpc), while the
middle and bottom rows include only \acc and \ins stars respectively.  Dashed
horizontal lines indicate the circularity cut used to define disc stars. Dashed
vertical lines mark the $5$~kpc cut in radius used to define `bulge' stars. All
stars outside these regions are classified as halo stars. The colour scale
corresponds to the logarithm of the number of star particles.}

\label{fig:circularityRadius}
\end{figure*}

\fig{fig:circularityRadius} shows the distribution of stellar circularity as a
function of radius in our three simulations. A concentration of corotating
stars on near-circular orbits extending to ${\sim30}$~kpc is obvious in all
cases, which we identify with the thin disc.  A number of streams on pro- and
retrograde orbits are also visible at large radii.

At $r\lesssim 5$~kpc the density of stars on non-circular orbits is comparable
to the density of stars in the disc. We identify this complex region with a
galactic `bulge' (\REF{using the term loosely, since this region is
substantially more extended than the bulge of the Milky Way}). To simplify our
definition of the stellar halo, we exclude \textit{all} stars with $r<5$~kpc,
regardless of circularity. This cut is easy to apply to both models and data.
It also follows the loose convention of most Milky Way stellar halo work, in
which stars more than a few kiloparsecs interior to the solar neighbourhood are
excluded (the exception being those high above the disc plane) even though the
inward extrapolation of a canonical $r^{-3}$ density profile would predict a
substantial mass of halo stars in the centre of the Galaxy (see also the
discussion in \citealt{Cooper10}). 

\fig{fig:circularityRadius} further separates star particles into \acc (middle
row) and \ins (bottom row) according to whether or not they are bound to the
main branch progenitor of each DM halo at the first snapshot after
their formation. Star particles that are first bound to a DM halo
other than the main progenitor are considered as \acc, even if they form in a
subhalo of the main branch (i.e. if they form in a satellite galaxy of the
Milky Way analogue) and are subsequently stripped\footnote{This is an important
difference with the work of \tthirteen{}, who included stars formed in bound
satellites within $r_{200}$ in the \ins halo as part of their `endo debris'
category.}.  ~\tab{tab:overallProps} summarizes the total mass of the stellar
halo and the relative proportion of \ins stars.

\begin{table}
\center
\label{tab:overallProps}

\caption{Total mass in the (outer) disc, `bulge' and stellar halo regions of
the $(r, \mathcal{E}_{\mathrm{E}})$ plane according to our criteria.  The final row
gives the fraction of mass in the stellar halo region that is formed \ins. Our
galaxies are roughly half the mass of the Milky Way; note that the disc mass
quoted is only for stars with $r>5$~kpc. Our `bulge' definition includes all
stars with $r<5$~kpc, regardless of their kinematics \REF{(it does not
correspond to a specific kinematic or photometric component in our simulations
or in real galaxies)}.} 

\begin{tabular}{clccc}
\hline 
\hline
 & & Aq-C & Aq-D & Aq-E \\ 
\hline
\multirow{3}{1cm}{Mass\\$(10^{9}\msol)$} & Disc  ($r>5$~kpc, $\mathcal{E}_{\rm E} > 0.8$)       &  3.0     &  4.3     & 4.3  \\
& Bulge  ($r<5$~kpc)                                   & 34.5     & 21.9     & 25.2 \\
& Halo ($r>5$~kpc, $\mathcal{E}_{\rm E} < 0.8$)&  4.6     &  8.4     & 6.8  \\
\hline
\multicolumn{2}{c}{\textit{In situ} halo fraction} &  37\%  &  33\%  & 41\% \\
\hline
\end{tabular}
\end{table}

\begin{figure}
  \centerline{\includegraphics[width=1.0\linewidth,trim=0.0cm 0cm 0cm 0cm, clip=True]{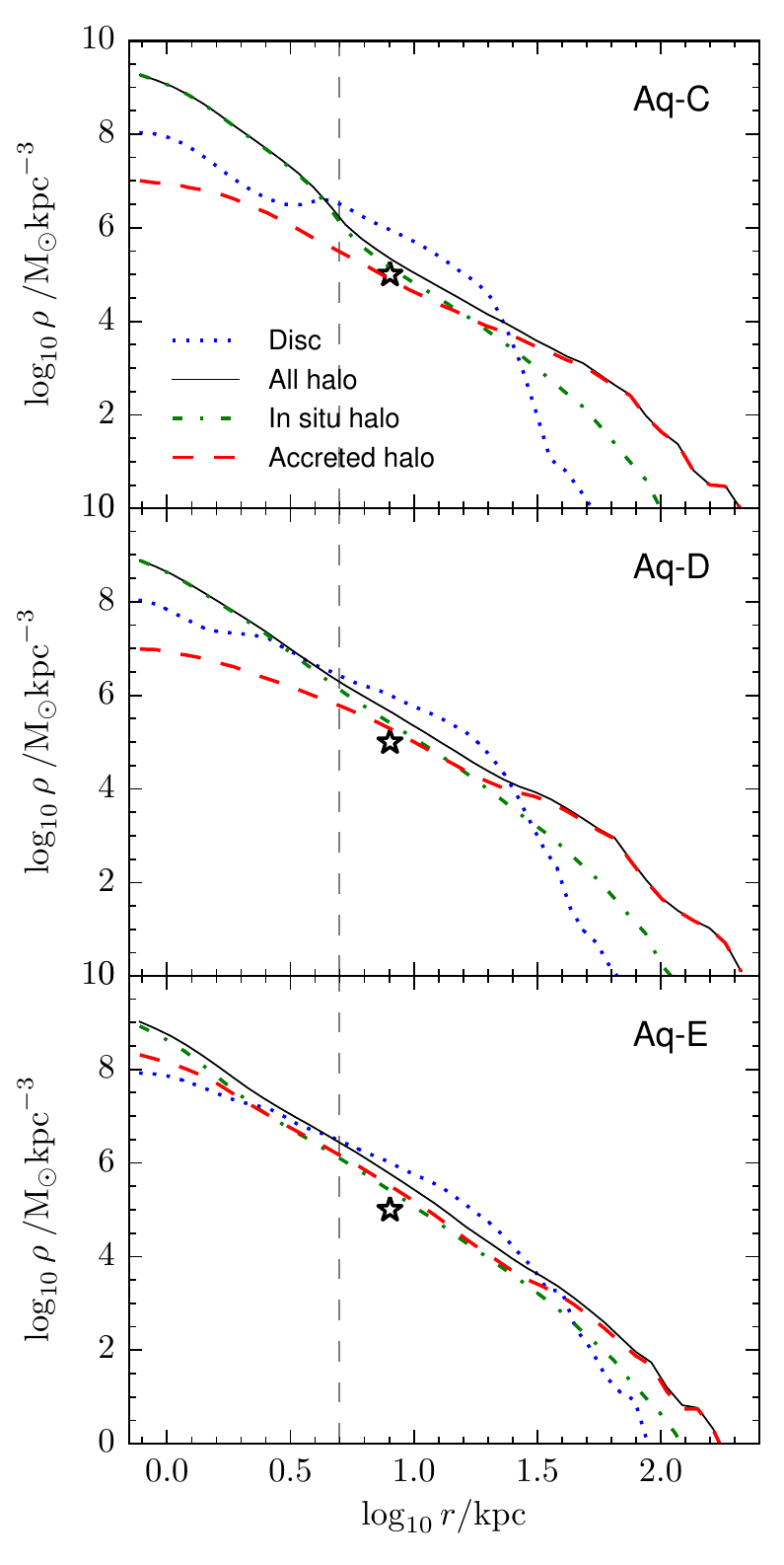}}

\caption{Spherically averaged stellar density profiles for \aqc (top), \aqd
(centre) and \aqe (bottom). The red dashed and green dot--dashed lines
correspond to the \acc and \ins halo components respectively. The solid black
line is the sum of these and can be compared to measurements for the Milky Way
\citep[black star;][]{FuchsJahreiss1998,Gould1998}. The blue dotted line
corresponds to disc stars, selected by their orbital circularity. The minimum
extent of the radial axis is set by the gravitational softening length for star
particles. A grey dashed vertical line at $5$~kpc marks the `bulge' region that
we exclude from our analysis of the stellar halo.} \label{fig:densityProfiles}
\end{figure}

\fig{fig:densityProfiles} shows the mean density of \acc and \ins halo stars in
spherical shells centred on the galaxy. The profile of the \ins halo has a
similar shape and amplitude in all three simulations, with a slight steepening
evident in the `bulge' region of \aqc. In both \aqc and \aqd, the \acc halo
stars are less centrally concentrated than the \ins component, with a mild
break due to \acc stars alone at $70<r<90$~kpc, while in \aqe the two
components are almost indistinguishable. The \acc--\ins transition in these
profiles at $\sim20$~kpc is consistent with the average for Milky Way analogues
in the GIMIC simulation \citep{Font2011b}.

\aqd and \aqe have `thick' discs with a high degree of non-circular motion,
apparent in the top row of \fig{fig:circularityRadius} as a high density of
stars at $0.5 < \mathcal{E}_{\rm E} < 0.8$ and $5 < r < 20$~kpc. According to
the aformentioned cuts on circularity and radius, we classify these as halo
stars. However, examining the variation of the circularity distribution with
height above the disc plane reveals that these stars simply make up the
low-circularity tail of a continuous distribution. The fraction of stars on
circular orbits is highest close to the plane.

The most important question \REF{from the point of view of} this paper is not
the origin of thick disc stars, but whether or not they can, or should, be
distinguished as a separate galactic component. Our simulations differ greatly
in this respect, in line with earlier studies of the origin and structure of
thick discs in simulations \citep[e.g.][]{Sales09}. \fig{fig:circularityRadius}
shows that \acc stars can make a significant contribution to the `disc'. In
\aqd, they contribute mainly to the `thin' disc -- the thick disc is formed
\ins. In \aqe, \acc and \ins `disc' stars contribute at a similar ratio over a
wide range of circularity and radius. 

\REF{In the context of observations of the Milky Way, a geometrically oblate
stellar component, intermediate between disc and halo, was identified by
\citet{Yoshii82} and \citet{Gilmore83}, but optimal and objective ways to
classify this component are still under debate \citep[see, for example, recent
reviews by][]{IvezicReview12,RixBovyReview13}.  Classifications have been
suggested based on metallicity and kinematics that imply different distinctions
between thick disc and halo stars \citep{Ivezic08, Bovy2012,
RixBovyReview13,Schlesinger14, Ruchti:2015aa}. The chemodynamical distribution functions of
our three galaxies are different from one another, and almost certainly
different from the Milky Way distribution function on which these observational
definitions are based. Therefore, we do not believe that additional cuts to
separate a thick disc component from the thin disc and halo (for example, using
a threshold metallicity, a limiting rotation velocity or vertical extent) are
helpful for the interpretation of our simulations.  This should
be kept in mind when comparing our results to observations of the Milky Way.}

\section{The Origin of \textit{In Situ} Stars} \label{sec:ORIGIN}

In this section we look in more depth at the origin of the \ins component of
the stellar halo. The density of this component exceeds that of \acc halo stars
in the inner $\sim20$ kpc of our galaxies. It may thus be very important for
spectroscopic observations of halo stars in the solar neighbourhood and in
surveys of main-sequence turnoff stars within a few kiloparsecs of the Milky
Way disc plane. 

In order to trace how \ins stars formed in our simulations, we separate them into three disjoint subcategories:

\begin{enumerate}
\item `heated disc' stars, which met the thin disc circularity criterion when they were formed, but are not in the disc at $z=0$;
\item stars formed from `stripped gas', brought into the main DM halo bound to a subhalo and subsequently stripped by tidal forces or ram pressure;
\item stars from `smoothly accreted gas', which enters the main DM halo through direct (smooth) accretion. 
\end{enumerate}

Categories (ii) and (iii) are easily distinguished by tracing the DM halo membership history of the parent gas particle for each star particle. The gas from which heated disc stars form must originally have either been stripped from a subhalo or smoothly accreted, so these stars could also be classified into the second or third categories. However, in this case it is the fact that they formed in the thin disc and were scattered out of it, rather than how their parent gas particle arrived in the disc, that we consider to be most important.

\begin{table}
\center
\caption{Breakdown of all \ins halo stars into three subtypes, according to their formation mechanism.
  \label{tab:insituBreakdown}}

\begin{tabular}{l c c c}\hline \hline
                         &  Aq-C     &  Aq-D      & Aq-E     \\ 
\hline
Heated disc              &  2.8\%    &  26.0\%    & 31.0\%   \\
Stripped gas             &  59.8\%   &  56.7\%    & 56.9\%   \\
Smoothly accreted gas    &  37.3\%   &  17.3\%    & 12.1\%   \\
\hline
\end{tabular}
\end{table}

\par

The fraction of \ins halo stars in each category is shown in \tab{tab:insituBreakdown}. It is clear that there is a large variation between the three simulations, although the \ins stars forming from gas stripped from satellites dominate in all cases.  In the next section we discuss each category in more detail and compare their properties.

\subsection{Heated disc stars}
 
The central galaxies in our three simulations undergo several episodes of disc
destruction and regrowth at $z > 3$. Over the redshift range $3 > z > 2$, a
stable disc is established. This disc continues to grow until $z=0$, although
its angular momentum axis may precess. Our heated disc category only includes
stars that once belonged to this stable disc. Stars on highly circular orbits
may be scattered to more eccentric orbits by secular evolution and satellite
impacts \citep[e.g.][]{Purcell2010}. We refer to this loosely as `heating', in
the sense of an increase in non-circular motion. These perturbed disc stars are
likely to have a clear kinematic and chemical relationship to those in the
present-day thin disc. 

We identify all star particles in the $z=0$ disc, as defined in
\sect{sec:SAMPLE}, that exist in a given earlier snapshot and use these to
define $J_{z}$ (assuming that the number of star particles scattered
\textit{into} the disc is negligible). We then apply the circularity threshold
$\mathcal{E}_{\rm E} > 0.8$ to identify all newly formed star particles in the disc
at that snapshot. Any of these that have $\mathcal{E}_{\rm E} < 0.8$ at $z=0$ are
assigned to our heated disc category. 

Beyond a certain redshift, $z_{\mathrm{form}}$, we can no longer reliably
identify a stable progenitor of the $z=0$ disc, and hence we cannot define
$J_{z}$. This limit is due to the small number ($<100$) of ancient disc stars
and increasing frequency of fluctuations in the central potential that
destabilize protodiscs.  ~\tab{tab:discFormation} gives $z_{\mathrm{form}}$ for
each of our simulations, along with $z_{1/2}$, the redshift by which the $z=0$
disc has assembled half its final mass. ~\tab{tab:discFormation} also gives
$f_{\mathrm{form}}$, the mass fraction of $z=0$ disc stars that form earlier
than $z_{\mathrm{form}}$.  This fraction is no more than 16 per cent (\aqe).
Stars scattered from this unidentified protodisc (and any others discs that
were completely destroyed before $z_{\mathrm{form}}$) are considered to fall
into one of the other two \ins categories, according to the origin of their
parent gas particle. 

\begin{table}
\center
\caption{Properties of the present-day thin stellar discs in our three simulations, defined by circularity $\mathcal{E}_{\rm E} > 0.8$. Columns from left to right give formation redshift, mass fraction in place at formation redshift, and redshift at which half the $z=0$ mass is in place.
  \label{tab:discFormation}}
  
\begin{tabular}{l c c c}\hline \hline
     & $z_{\mathrm{form}}$ &  $f_{\mathrm{form}}$ &  $z_{\mathrm{1/2}}$ \\
\hline
Aq-C &2.54 & 12\%  & 0.92 \\ 
Aq-D &2.32 &  8\%  & 0.83 \\ 
Aq-E &2.20 & 16\%  & 0.76 \\ 
\hline
\end{tabular}

\end{table}

\subsection{Stars from stripped gas and smoothly accreted gas}
\label{sec:STRIPPEDSMOOTH}

Halo stars can form directly in the circumgalactic medium, either in quasi-free-falling cold gas clouds (not associated with DM clumps) or the gaseous tidal or ram pressure stripped streams of satellite galaxies.  We distinguish between these two possibilities based on whether or not the parent gas particle of a given star particle was bound to another DM halo before being bound to the main halo. Stars forming from stripped satellite gas particles may be chemically and kinematically similar to stars in the \acc stellar halo. In contrast, stars forming in gas condensing out of the hot hydrostatic gas halo, or other `smoothly' accreted cold clumps, may have properties more similar to those expected of an \ins halo formed by monolithic collapse.

\begin{figure}
  \centerline{\includegraphics[width=1.0\linewidth,trim=0.0cm 0.5cm 0cm 0.5cm, clip=True]{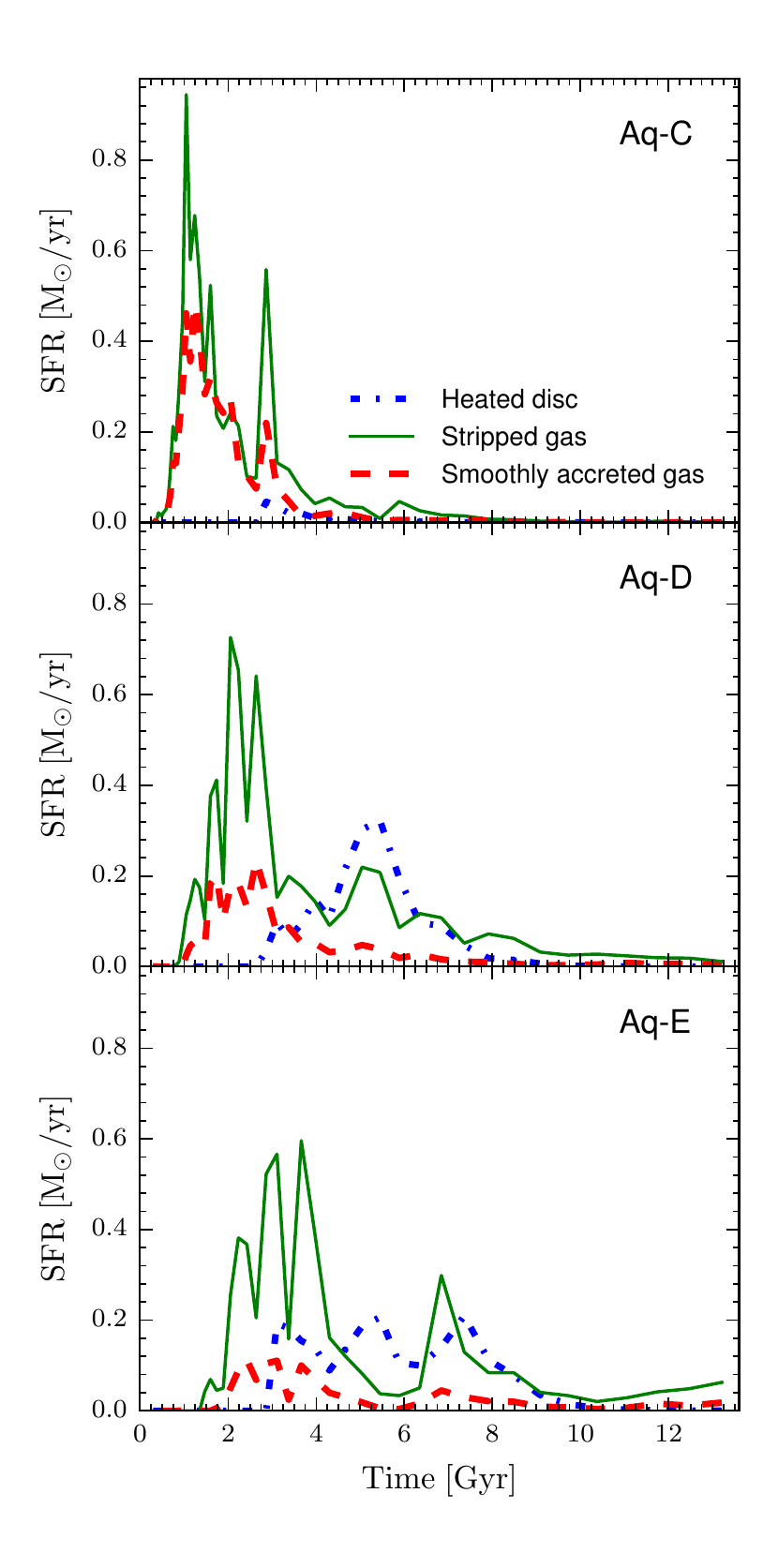}}
\caption{Formation history for \ins halo stars assigned to each of our three \ins formation mechanisms. Time is measured from the origin of the universe. Peaks in the formation of stars in smoothly accreted gas are clearly correlated with those in stripped gas at $t<4$~Gyr. Vertical dotted lines mark $z_{\mathrm{form}}$ and $z_{1/2}$ as given in ~\tab{tab:discFormation}.}
\label{fig:sfHistory}
\end{figure}

\fig{fig:sfHistory} shows the absolute star formation rate of each \ins
category as a function of time elapsed since the big bang. These star formation
rates are low compared to those typical of the stable disc and the progenitors
of \acc stars ($\sim 1 \, \msol \, \mathrm{yr^{-1}}$). The majority of halo
stars that form in stripped or smoothly accreted gas are more than 9~Gyr old,
only marginally younger than the typical age of \acc stars. In \aqd and \aqe,
there are also $\sim2$~Gyr-long bursts of \ins star formation \REFB{starting at 
$t\approx4$ and $6$~Gyr respectively}. Interestingly, these also correspond to
episodes of formation for scattered disc stars (blue) and thin disc stars (not
shown). \REF{These episodes correspond to the rapid infall of cold gas on to the
disc during periods in which several relatively massive satellites are being
disrupted simultaneously.}

Another notable feature of \fig{fig:sfHistory} is that the star formation rate
in smoothly accreted gas (red) is clearly correlated with that in stripped gas
(green), especially in the dominant early epoch of \ins halo formation (ages
$>9$~Gyr). This correlation persists even if we select only stars forming at $r
> 30$~kpc, far away from the disc, suggesting that the conditions under which
\textit{most} \ins halo stars form are in fact related to the accretion and
stripping of gas-rich satellites.  It appears that star formation may be
triggered by the mixing of free-floating gas from the hydrostatic halo with
star-forming stripped gas. We also see corresponding peaks in the \acc halo
star formation rate, suggesting that star formation is triggered in the
infalling satellites as well. 

\REFB{These effects must arise from the subgrid model of star formation. They
may therefore occur, to a greater or lesser extent, with other similar star
formation prescriptions used in the literature.  Since there are no unambiguous
observational tests of the predictions of models for star formation in the
diffuse circumgalactic gas, it is very difficult to judge whether or not the
behaviour of a particular simulation is physically plausible (see also the
discussion in Appendix B).  Isolating dependencies on modelling uncertainties
such as these is one of our motivations for separating simulated halo stars
into categories based on their physical origin.}

\REFB{Regardless of how they actually form, the \textit{classification} of gas
particles as `smoothly' accreted is resolution dependent: gas bound to a
low-mass infalling DM halo, and so classified as `stripped' at high
resolution, would  be classified as `smoothly accreted' in a lower resolution
simulation that does not resolve the parent halo. In Appendix B we conclude
that, in our simulations, the uncertainty in the total mass of the smoothly
accreted halo component is dominated by this classification uncertainty, rather
than the resolution dependence of star formation efficiency.}

\section{In situ haloes at \MakeLowercase{$z=0$}}
\label{sec:PROPTODAY}

In this section, we examine the observable characteristics of \ins halo stars at the present day, starting with a summary of halo properties and then looking in more detail at regions analogous to the solar neighbourhood.

\subsection{Whole halo}

\begin{figure}
  \centerline{\includegraphics[width=1.0\linewidth,trim = 0.0cm 0.0cm 0cm 0cm, clip=True]{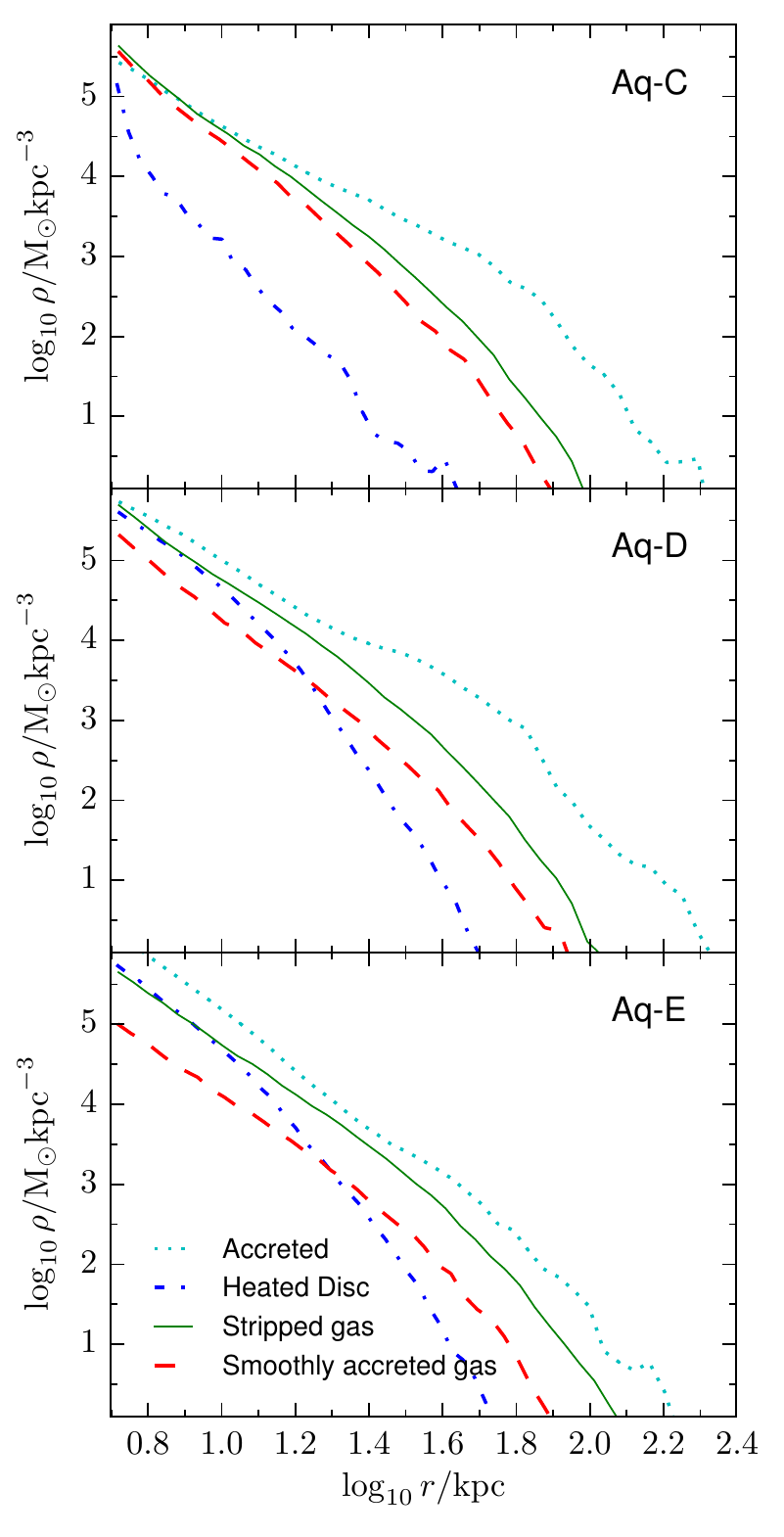}}
\caption{Spherically averaged density profiles of halo stars in the \acc and three \ins components for \aqc (top), \aqd (centre) and \aqe (bottom).}
\label{fig:componentDensityProfile}
\end{figure}

\fig{fig:componentDensityProfile} compares the spherically averaged density
profiles of our three \ins halo categories and \acc halo stars. In the
$r<20$~kpc region where \ins halo stars dominate over \acc stars, they
contribute roughly equal mass fractions; the exact proportions vary from halo
to halo. We see a strong correspondence between stars formed from stripped and
smoothly accreted gas at all radii, which, in combination with
\fig{fig:sfHistory}, suggests that they form with a similar distribution in
both space and time. As expected, heated disc stars have a steeper profile,
with most concentrated at $r<20$~kpc.   

\begin{figure} \centerline{\includegraphics[width=1.0\linewidth, trim = 0.1cm 0.1cm 0cm 0cm, clip=True]{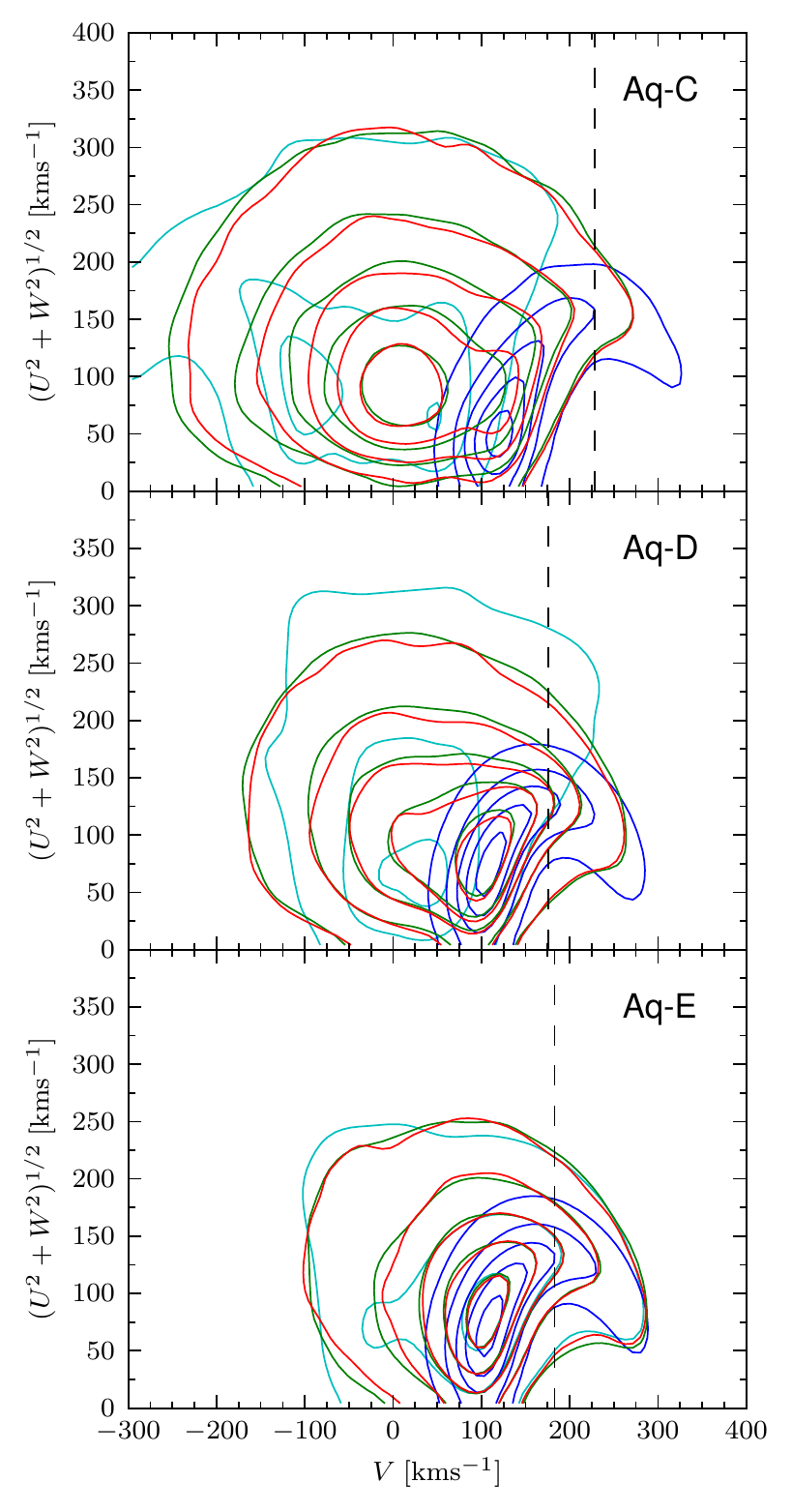}}
  
  \caption{Toomre diagrams of the whole
  stellar halo. For stripped-gas, smooth-gas and heated disc \ins halo stars
  (red, green and dark blue respectively), contours mark the regions enclosing
  10, 30, 50, 70 and 90 per cent of the stellar mass. For the \acc halo (cyan)
  only 10, 50 and 90 per cent levels are shown.  The dashed vertical line marks
  the rotation velocity of the disc at 8~kpc.} 
  
  \label{fig:overallToomre}
\end{figure}

\fig{fig:overallToomre} shows Toomre diagrams \citep{Sandage87} that compare the amplitude of circular and radial motion for different components. A galactocentric $UVW$ velocity frame \citep[e.g.][p. 627]{BM98} is defined with respect to the thin disc in each simulation. 

Of the three simulations, the stellar halo in \aqc has kinematic properties most similar to those  measured for the Milky Way. The peak rotational velocity of the disc is $\sim220 \, \mathrm{km \, s^{-1}}$. The heated disc stars (blue) rotate in the same sense as those on circular orbits, with a lag of $\sim40$--$180 \kms$. Stars formed from stripped and smoothly accreted gas are kinematically indistinguishable from each other, once again pointing to a close correlation between the dynamics of the two components. In the Toomre diagram they resemble the classical Milky Way halo, with zero net rotation and high radial velocity dispersion. Accreted halo stars show a similar distribution overall, but with notable overdensities due to individual streams, some of which have a net retrograde motion. 

The heated disc stars in \aqd and \aqe have similar kinematics to those in \aqc, but the stripped/smooth \ins haloes have a greater net rotation. In \aqe, all three components once again resemble one another, although the stripped- and smooth-gas halo stars have a greater velocity dispersion. An underlying stripped/smooth \ins halo may still be present, but the bulk of \ins halo stars are more similar kinematically to the Milky Way thick disc. The behaviour of \acc stars once again resembles that of the \ins component, even to the extent that they have a strong prograde rotation in \aqe. Accreted halo components with prograde rotation were noted by \citet{Abadi2003} and also found in Milky Way-like systems in the GIMIC simulations \citep{Font2011b, McCarthy12}

\begin{figure}
  \centerline{\includegraphics[width=1.0\linewidth,clip=True,trim=0.0cm 0cm 0cm 0cm]{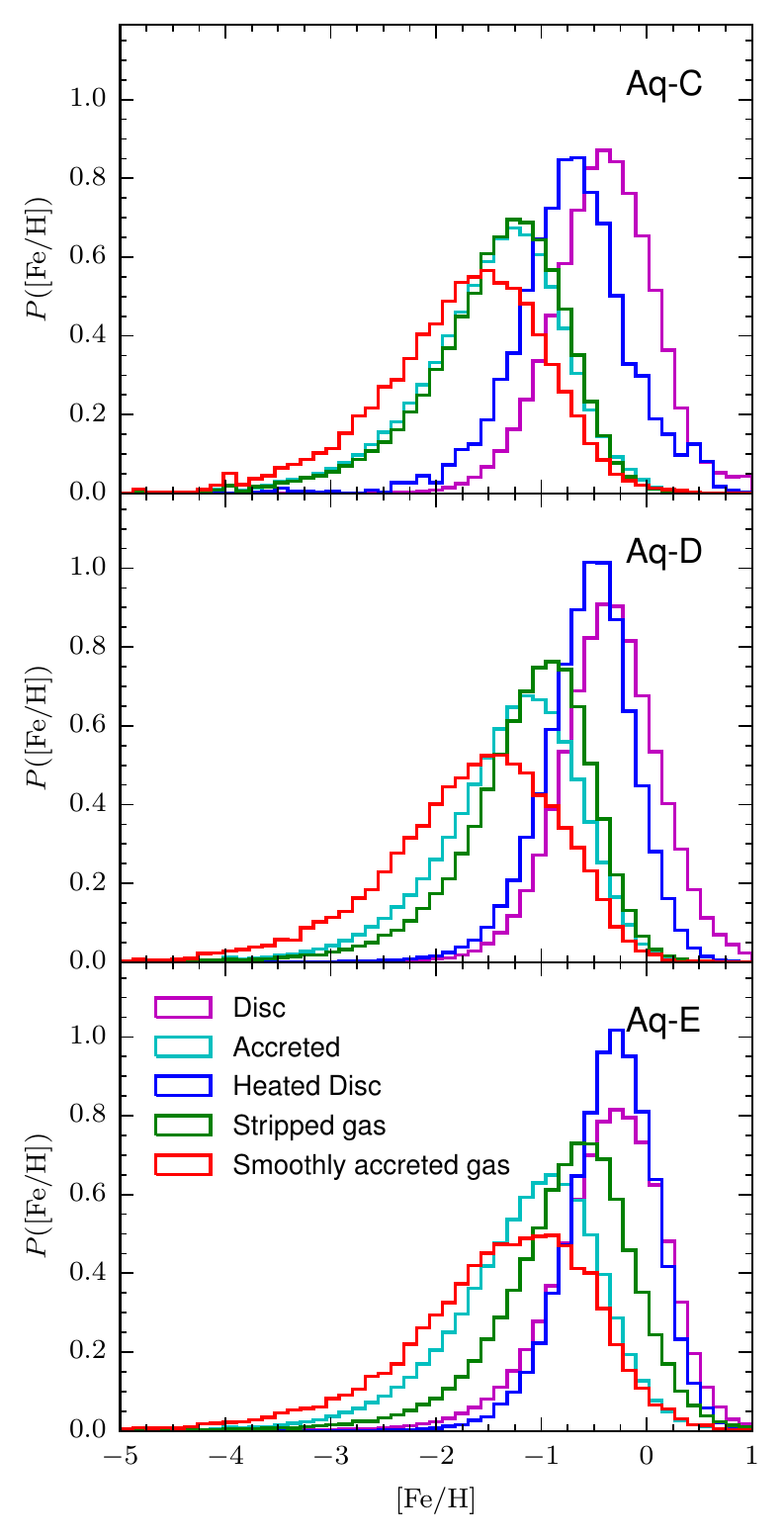}}
\caption{[Fe/H] distributions for the disc, \acc halo and three \ins halo components. Distributions are normalized by the total mass of stars in each component.}
\label{fig:overallFeH}
\end{figure}

Finally, in \fig{fig:overallFeH}, we examine the normalized metallicity distribution functions (MDFs) of each component of the \ins halo.  Heated disc stars have the highest median [Fe/H] and narrowest dispersion. Their MDF resembles that of the thin disc, but is slightly more metal poor (by $\sim0.5$~dex in \aqc). Both \ins and \acc halo stars are systematically more metal poor than heated disc stars. 

The MDF of \ins halo stars formed from stripped gas is very similar to that of \acc satellite stars, with a median systematically higher by no more than 0.1~dex. This is to be expected, as the dense cold gas stripped from satellites will have been enriched by the same stellar populations that make up the \acc halo. Moreover, very similar distributions will also result if prolonged star formation occurs in satellite galaxies while their gas is being stripped. The overall \ins MDF is close to that of the stripped-gas stars, since they dominate the \ins mass budget.

Looking in detail, the degree of similarity between the MDFs of the various
components varies in each of our three simulations. This may depend on the
extent to which the satellite galaxies contributing the bulk of stripped-gas
stars are the same as those that contribute the majority of \acc stars.
\REF{Since} gas can be more easily expelled from shallower potentials, the most
massive and metal-rich accreted progenitor galaxies are likely to retain the
most gas when they enter the main DM halo. Stars stripped from these
galaxies are expected to dominate the \acc halo, particularly near the centre.
\REF{We investigate the relative contributions of star-forming stripped gas and
directly accreted stars from different progenitors in
Section~\ref{sec:PROGENITORS} below.}

Of the different \ins components, it is the stars that formed from smoothly
accreted gas that have the lowest median metallicity and the broadest
dispersion. This is consistent with the expectation that the gas surrounding each
galaxy will be a mix of its own metal-rich ejecta and a large quantity of
`pristine', or only marginally enriched, gas from direct cosmological infall
\citep{Crain2010a}. In our simulations, the MDF is the only clear distinction
between `smooth gas' stars and `stripped gas' halo stars.

\subsection{The Solar Neighbourhood}
As a rough analogue of the solar neighbourhood region most relevant to current observations, we examine the average properties of halo stars in a torus of cross-sectional diameter 4~kpc and galactocentric radius $r=8$~kpc in the plane of the thin disc.

~\tab{tab:insituBreakdown_sn} summarizes the fraction of stars in each component. For a more direct comparison to the real data, we have grouped heated disc stars and stars that meet our thin disc circularity cut into a single disc component, because the typically high circular velocities of heated disc stars would most likely result in them being classified as `thick disc' rather than halo stars in observations.  Approximately 10 per cent of the stellar mass then remains in a component resembling the `classic' halo, of which \acc stars contribute between 34 and 67 per cent. 

\begin{table*}
\center
\caption{Breakdown of all stars in the solar neighbourhood. Heated disc stars are grouped together with thin disc stars in this table. The top two rows give fractions of total stellar mass, while the lower three rows give fractions of stellar halo mass (second row) only.
\label{tab:insituBreakdown_sn}}

\begin{tabular}{l c c c}\hline \hline
Mass  $(10^{8}\,\msol)$ & Aq-C & Aq-D & Aq-E \\ \hline
Disc stars (thin + thick)              & 26.4  (92.3\%) & 27.4  (88.7\%)  & 28.0  (83.0\%) \\
Halo stars                             & 2.21  (7.7\%)  & 3.49  (11.3\%)  & 5.72  (17.0\%) \\
\hline
\hspace*{0.5cm} Accreted               & 0.747 (33.9\%) & 1.79  (51.1\%)  & 3.83  (67.0\%) \\
\hspace*{0.5cm} Stripped gas           & 0.837 (37.9\%) & 1.23  (35.3\%)  & 1.53  (26.7\%) \\
\hspace*{0.5cm} Smoothly accreted gas  & 0.618 (28.0\%) & 0.476 (13.6\%)  & 0.356 (6.23\%) \\
\hline
\end{tabular}

\end{table*}

\begin{figure}
  \centerline{\includegraphics[width=1.0\linewidth,clip=True,trim=0.0cm 0cm 0cm 0cm]{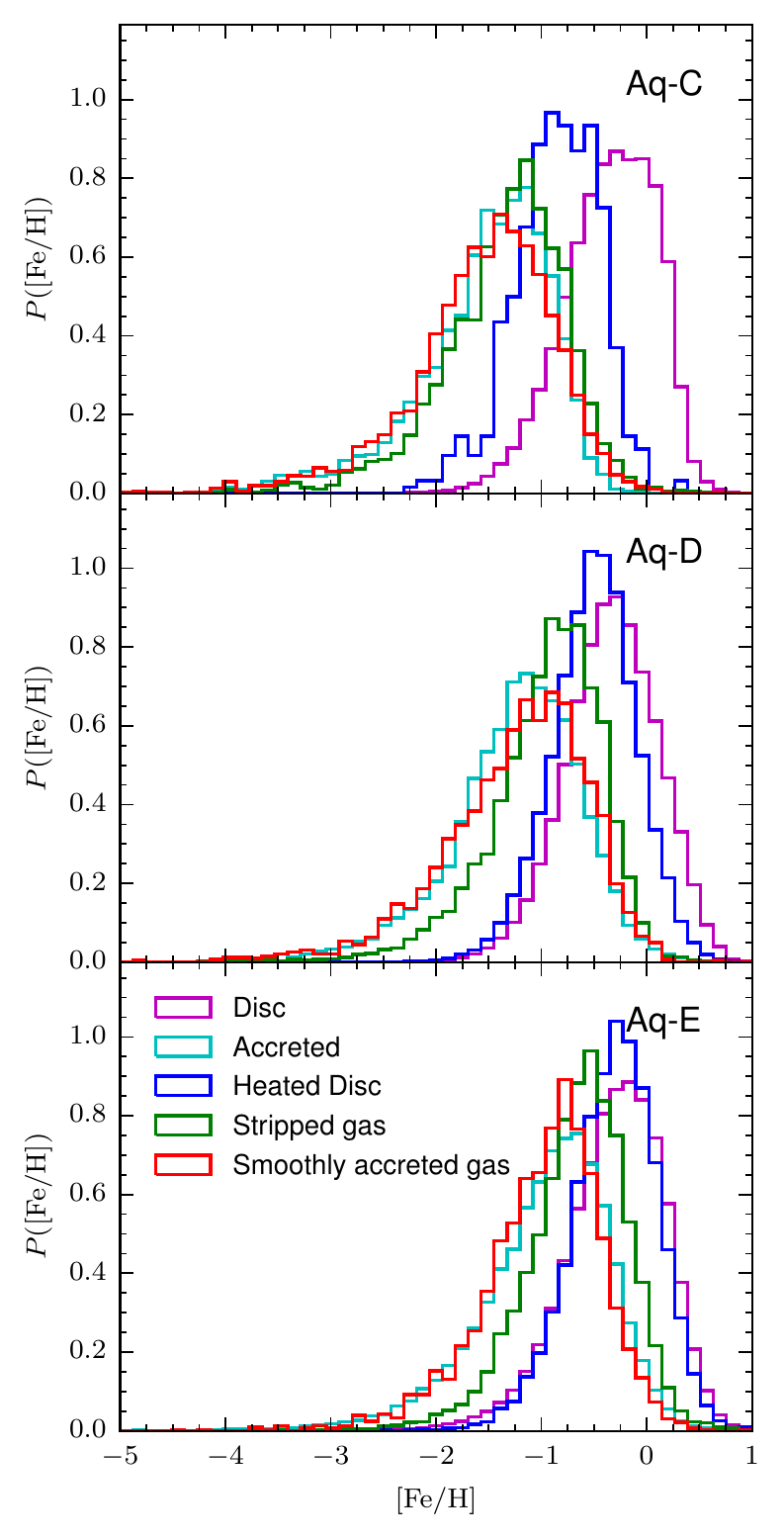}}
\caption{MDFs, as Fig.~\ref{fig:overallFeH}, but here in the solar neighbourhood.}
\label{fig:solarNeighbourhoodFeH}
\end{figure}

Toomre diagrams in this region are almost identical to those in \fig{fig:overallToomre}. The biggest differences in comparison to the overall halo are found in the solar neighbourhood MDFs, which are shown in \fig{fig:solarNeighbourhoodFeH}. Stars formed from smoothly accreted gas that end up in the solar neighbourhood are more metal rich on average, such that their MDF has a very similar shape and amplitude to the \acc halo. This may be because the metal-poor contribution of this component seen in \fig{fig:overallFeH} is dominated by stars forming at large radii from gas that has not been polluted by the galactic wind of the central galaxy. Other components have the same relationship to one another as those in \fig{fig:overallFeH}. Hence, we find no substantial differences between the properties of the \ins halo in the solar neighbourhood and the \ins halo overall. This is not surprising because we have already seen that the bulk of the \ins halo is concentrated within $r\lesssim20$~kpc.

\section{Satellite Progenitors} \label{sec:PROGENITORS}

We have shown that our simulated stellar haloes are dominated by satellite accretion: the bulk of halo stars are stripped directly from satellites, and the majority of `\ins' halo stars form from gas stripped from satellites. However, as \fig{fig:densityProfiles} demonstrates, stars formed \ins from stripped gas have a more centrally concentrated spatial distribution at $z=0$ than directly accreted stars. In this section we examine the satellites which contribute to the halo, with the aim of determining how their infall times, masses and baryonic content affect the spatial distribution of the \ins and \acc components.

\begin{figure}
  \centerline{\includegraphics[width=1.0\linewidth]{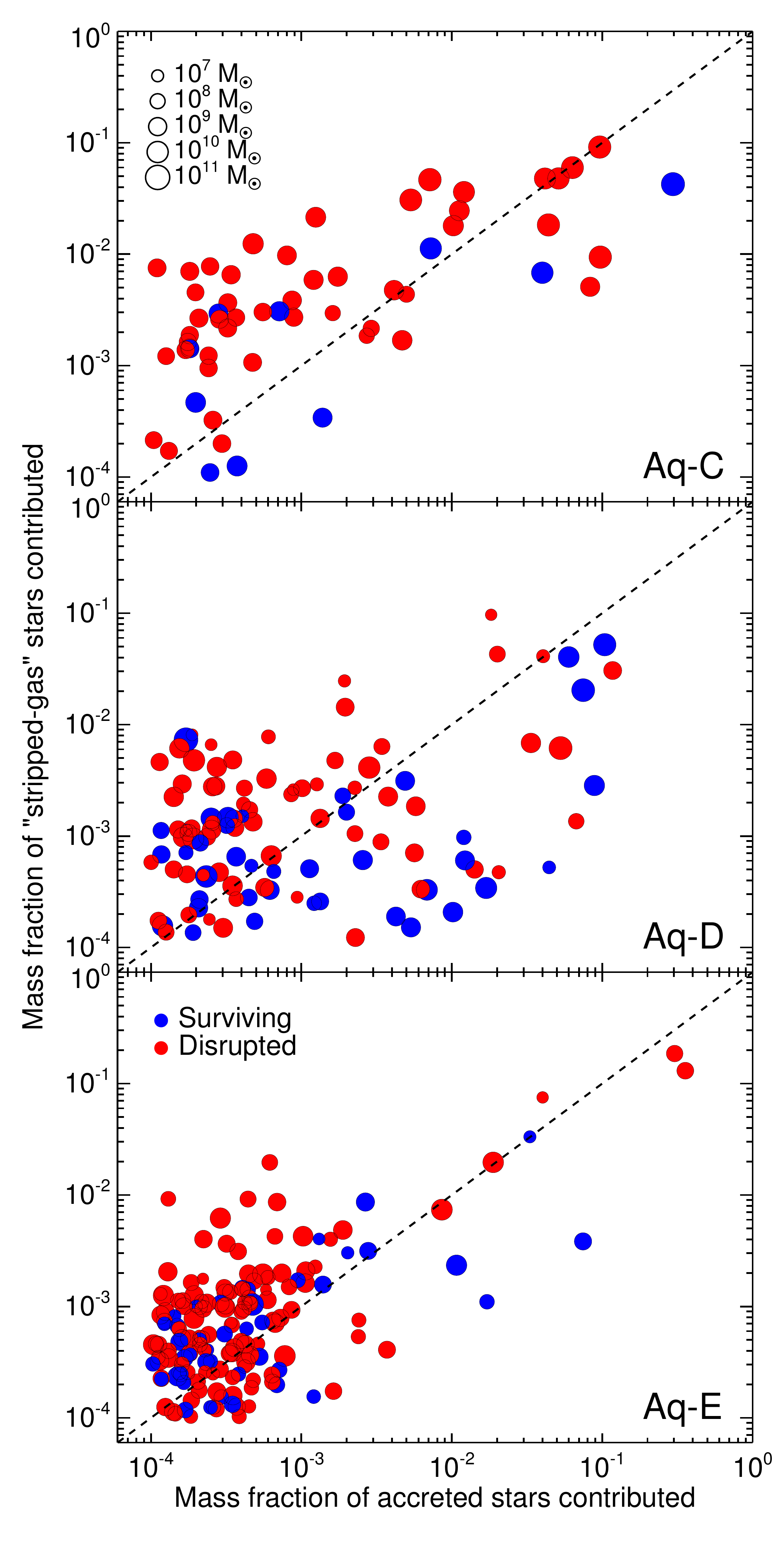}}
\caption{Mass fraction of the \acc halo and stripped-gas halo contributed by disrupted (red) and surviving (blue) satellite galaxies.  The sizes of the points are proportional to the logarithm of the satellite's total mass at infall, as shown by the legend in the first panel.  The diagonal dashed line indicates an equal fractional contribution to the \acc and stripped-gas components. }
\label{fig:acc_str_contrib}
\end{figure}

We first ask whether the subset of satellite progenitors contributing the gas
from which an \ins stellar halo forms is the same subset contributing accreted
stars.  \fig{fig:acc_str_contrib} compares the mass fractions of stars formed
from stripped gas and directly accreted stars associated with each progenitor
satellite.  In all three simulations, satellites that contribute significantly
to one component also tend to contribute significantly to the other.  There is
substantial variation in detail between the three haloes, reflecting their
different accretion histories. A larger scatter is apparent in \aqd, as well as
a noticeable fraction of gas-poor contributors (lower-right area of the plot)
relative to \aqc and \aqe.  A larger fraction of those satellites also survive
to $z=0$ without being disrupted (blue points).  The smaller number of
surviving satellites in \aqc reflects a quieter recent merger history.

\begin{figure}
  \centerline{\includegraphics[width=1.0\linewidth]{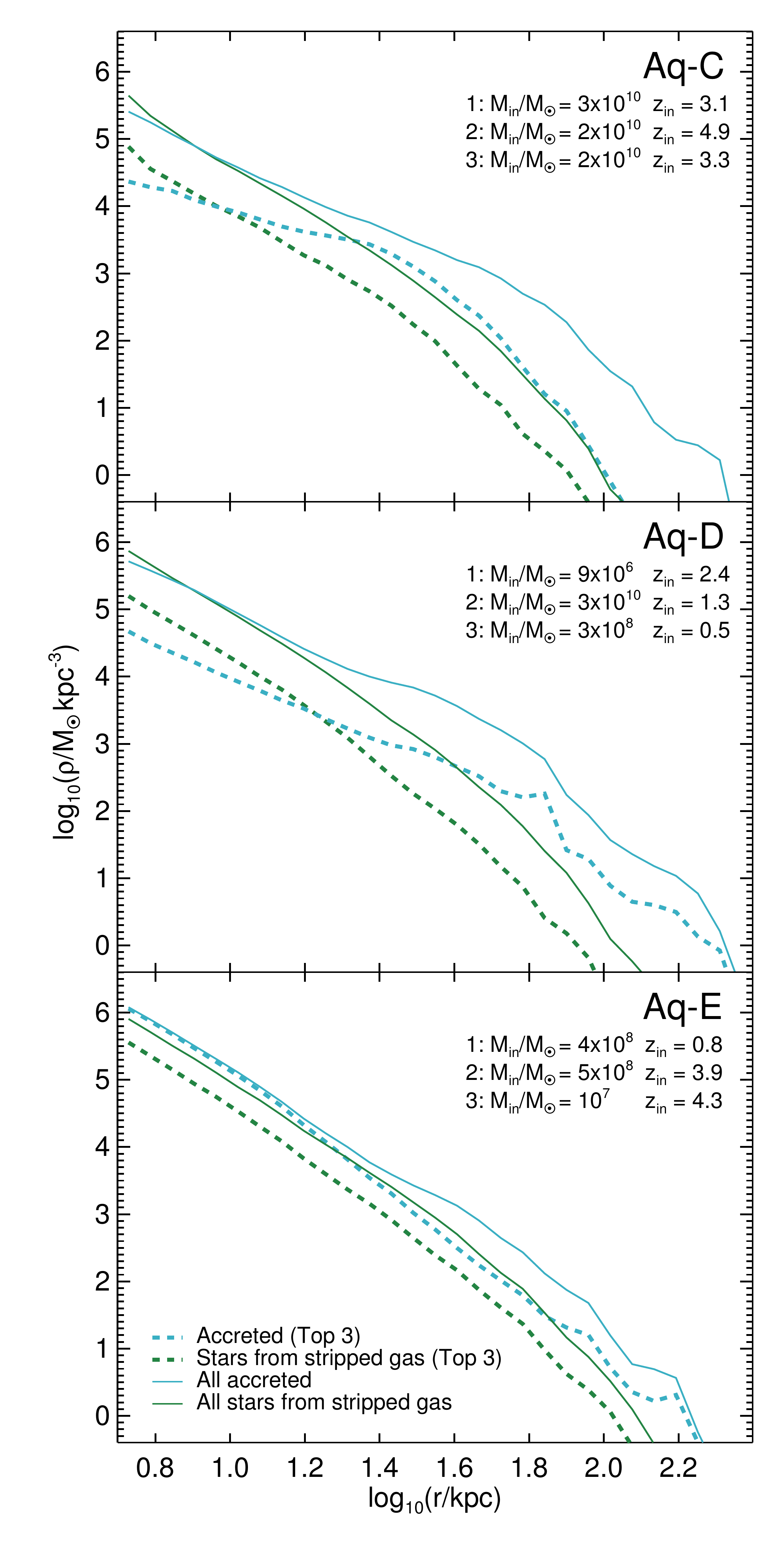}}
\caption{Spherically averaged density profiles of halo stars associated with the top three contributors to the `stripped-gas' component.  The profiles of stars formed from gas contributed by those galaxies are shown as dashed, cyan lines, while the profile of stars accreted from those satellites directly are shown as solid, green lines.  Thinner, black lines of the same styles show the corresponding total profiles.}
\label{fig:top_contrib_dens_prof}
\end{figure}

In \fig{fig:top_contrib_dens_prof} we isolate the top three satellite
progenitors of the stripped-gas \ins component and plot the density profiles of
the accreted and stripped-gas stars they contribute. \aqe stands out as having
the most similar profiles for the two components, both of which are slightly
steeper than the total accreted profile.  In this case, the satellites plotted
account for around 40 per cent of the total stripped-gas halo. With the
possible exception of accreted stars in Aq-C, both accreted and stripped-gas
stars from the top three satellites are distributed like the bulk of the
stellar halo.

\fig{fig:top_contrib_dens_prof} suggests that the greater central
concentration of the stripped-gas halo profile is not simply because most of
the progenitor gas particles originate in more massive progenitors, which sink
more rapidly through the action of dynamical friction. If that were the case, we
might also expect stars accreted from the same progenitors to be more centrally
concentrated than the accreted halo overall. Instead, we see that the debris
profiles of the three most massive individual progenitors are very similar to
the profile of the entire accreted halo. In our simulations, at the present
day, accreted stars are distributed (on average) over the same range of radii
at which they were liberated from their parent satellites. Conversely, we have
confirmed that the stripped-gas particles from which \ins halo stars form
dissipate some of their orbital energy between the times of stripping and star
formation. The present-day distribution of the stripped gas halo is therefore
imprinted at the time those stars form, rather than the time at which their
parent gas particles are stripped.

\section{Discussion} \label{sec:DISCUSSION}

Our finding of accretion-triggered star formation in smoothly accreted gas may
be, at least in part, a consequence of our hydrodynamics scheme. In some
hydrodynamic models, cold clouds can condense independently of tidally stripped
gas, while in others, condensation and mixing by fluid instabilities may both
be suppressed \citep[e.g.][]{Kaufmann06,Keres12,Parry2012,Hobbs13}. \REFB{As we
noted in Section~\ref{sec:STRIPPEDSMOOTH}, there are no strong observational
test of different models at this level of detail. In Appendix~B, we carry out a
simple resolution study of the smoothly accreted gas component. At lower
resolution, more baryons are retained by the main halo and relatively more of
these baryons are classified as smoothly accreted.  The overall efficiency with
which smoothly accreted gas is converted to stars also varies systematically
with resolution, although (in our simulations) this is a minor contribution to
the change of \ins stellar halo mass, relative to the first two effects}. 

\REFB{Given such uncertainties}, it is worth noting that, with a single SPH
code, the different cosmological initial conditions of our three simulations
result in markedly different \ins halo properties at $z=0$. It is obvious that
the mass spectrum, arrival time distribution and prior star formation histories
of accreted satellites directly determine the properties of the \acc halo.
Remarkably, however, in our SPH simulations, these factors are also extremely
important in the formation of the \ins halo. Without satellite accretion, there
would be hardly any \ins halo stars in our simulations.  Moreover, it is
unlikely that all the stars we classify as heated disc halo stars would be
considered as such by observers. The vast majority are on orbits of moderate
circular velocity, corotating and coplanar with the thin disc, and would likely
be classified as `thick disc' stars as discussed in Section~\ref{sec:SAMPLE}.
In \aqc, a large fraction of these stars belong to the end of a bar in the
galactic plane that extends slightly beyond our `bulge' cut at 5~kpc. 

Our model is clearly not unique. Nonetheless, our \ins haloes are compatible with the basic observable properties of the Milky Way's stellar halo, if we exclude scattered disc  stars. Stars formed from stripped and smoothly accreted gas in the main halo are old and metal poor, with a significant tail to low metallicity. The kinematic and chemical similarities are closest in \aqc, which also has the most Milky Way-like thin disc.

Two of our haloes conform to the expectation from earlier work that \ins halo stars are more concentrated than \acc stars, but in the third, the \ins and \acc halo density profiles are almost identical. Interestingly, it is only in this third case that there is a significant difference in the MDFs of \ins and \acc stars. 

\citet{McCarthy12} studied the origin of \ins halo stars in 412 Milky Way-mass
galaxies from the GIMIC simulations, which have a DM particle mass
$\sim300$ times larger than our simulations. With the same definition of disc
stars and \acc stars, they find \ins halo mass fractions ranging from 20 to 60
per cent. Their \ins stars are, on average, younger and more metal rich than
their \acc halo stars, with a centrally concentrated oblate distribution and
prograde rotation. They demonstrate that these properties result from star
formation at $z\sim1$ in `proto-discs' that are subsequently destroyed. 

Although our three simulations produce total \ins mass fractions consistent
with the distribution found by \citet{McCarthy12}, there are some notable
differences in the origin of \ins stars. Accreted gas is the dominant
contributor to our \ins haloes, whereas \citeauthor{McCarthy12} find
that approximately half of their \ins halo mass is formed from gas that is
shock-heated to the virial temperature of the main DM halo. Our \ins
DM haloes form at somewhat higher redshift and we can trace back our
$z=0$ discs to $z\sim2.5$, supporting the suggestion of \citeauthor{McCarthy12}
that the Aquarius simulations have quieter-than-average accretion histories. As
\citeauthor{McCarthy12} do not compute the fraction of gas forming their \ins
component that was previously bound to satellites, it is likely that their
proto-disc stars mix together both stripped and smoothly accreted gas, and
possibly some fraction of our heated disc population. \citet{Font2011b} note
that formation from stripped gas is apparent in the same set of galaxies from
the GIMIC simulations and is likely responsible for their \ins stars at $r >
20$~kpc. We find that this mode of formation dominates at somewhat smaller
radii.  This may be a consequence of the better spatial resolution in our
simulations; a smaller gravitational softening length means satellite gas tends
to be more tightly bound, allowing it to sink further into the potential before
being stripped by ram pressure or tidal forces \citep[see also][]{Parry2012}.

As noted in the introduction, a series of papers by \tiss{} (T12, T13, T14)
examine \ins halo stars in six SPH simulations from the Aquila suite
\citep{Aquila}. They also use \gadgetthree for these simulations, but with
different implementations of subgrid physics and slightly lower resolution (a
factor of $5$ in DM particle mass; \citealt{Aquila}). Two of their simulations,
Aq-C and Aq-D, have the same initial conditions as the simulations with the
corresponding labels in this paper. To aid the reader in comparing our results
with this series of papers, we comment here on the differences between our
sample definitions and those of \tiss{}

T12 separated disc and halo stars with a circularity cut, $J_{z}/J_{\rm
circ}(E) > 0.65$, which classifies many more stars as `disc' than our cut.
Also, they did not restrict the height of disc stars above the mid-plane. They
excluded bulge stars with an energy criterion that roughly equates to a radius
of 5~kpc, comparable to our `bulge' cut (see T12). 

In the analysis of \tiss{}, \ins halo stars are divided into
`inner' (more bound) and `outer' (less bound) populations according to a cut in
relative binding energy. This equates to a radial cut in the range
$14<r<36$~kpc (15 and 19~kpc for Aq-C and Aq-D respectively). They further
subdivided these populations by origin, according to whether the stars were
formed outside the virial radius (`debris') or inside (`endo-debris'). This
definition is different from ours, which counts a present-day halo star particle
as `accreted' if it formed bound to a satellite subhalo, regardless of whether
the satellite was inside or outside the virial radius of the main DM
halo at the time. This may not be significant, because the fraction of \acc
halo stars formed in bound satellites after their infall is typically small
(see \citealt{McCarthy12}). 

The definition of heated disc halo stars in T13 requires the star particle to
have $J_{z}/J_{\rm circ}(E) > 0.65$ at its formation.  Our definition also
requires those stars to have been scattered from the disc that survives at
$z=0$; we would classify some fraction of their heated disc stars as formed
from either stripped or smoothly accreted gas. This may explain why T13 find a
much larger heated disc fraction in Aq-C despite their less stringent disc
circularity cut (indeed, they note that heated disc stars have retrograde
rotation in one of their simulations). 

Allowing for these differences of definition between our study and that of T13,
a number of similarities are clear. \REFB{The dominance of \ins over \acc halo
stars in our simulations at $r \lesssim 20~\kpc$ is also seen in their `inner
halo' populations}. Moreover, they also find that scattered disc stars make a
negligible contribution outside this region.  The kinematic properties, as
shown in our \fig{fig:overallToomre}, are similar. In Aq-C and Aq-D, T13 report
a median $\mathrm{[Fe/H]}$ of endo-debris stars lower than those of debris
stars by $0.17$ and $0.26$~dex, respectively.  This is the opposite of what
we find, but could be easily explained if the T13 endo-debris definition
includes a similar or larger fraction of stars formed from smoothly accreted
gas. \REF{This is supported, at least qualitatively, by our findings in
Appendix~\ref{appendix_tissera}, which indicate first that the star formation
efficiency for the central galaxy is much higher in the \tiss{} simulations,
and secondly that stars in our simulations meeting the \tiss{} endo-debris
definition are predominantly formed in satellites \textit{after} their
accretion by the main halo.}

\REF{Recently, \citet{Pillepich15a} described the \ins stellar halo of the
\eris{} simulation (a zoom of an isolated halo, $M_{200}=8\times10^{11}\msol$,
forming a late-type galaxy with $M_{\star}=3.9\times10^{10}\msol$;
\citealt{Guedes2011}). The particle mass of \eris{} is $\sim1/3$ of that in our
simulations.  They find a transition in the density profile between \ins stars
(including the disc) and accreted stars at $\sim10\kpc$.  Having cut out a
cylinder of height $\pm5\kpc$ and radius $15\kpc$ to excise the disc and bulge,
they define an inner halo (all stars $r<20\kpc$ excluding this cylinder) and an
outer halo (all stars $20<r<235\kpc$). They find that $\sim25$ per cent of stars in
their inner halo were formed \ins, falling to only $3$ per cent in their
outer halo. Applying the same cuts to our haloes C, D and E, we find \ins
fractions of $(83, 78, 76)$ per cent for an equivalent inner halo region and
$(29, 31, 65)$ per cent for an equivalent outer halo region.  These much larger
fractions are most likely because  \citeauthor{Pillepich15a} count both disc
and halo stars in their definition of \ins stellar mass -- our discs are
larger than the disc of \eris{} and hence are not excised completely by their
geometric cut.  If we count only our halo stars in these regions (excluding
disc stars by their circularity as above), the \ins mass fractions for our
three simulations are $(20, 18, 15)$ percent in the \citeauthor{Pillepich15a}
inner halo and $(14,11,17)$ per cent in the outer halo.  Larger `outer halo' \textit{in
situ} mass fractions in our simulations may well result from differences between
our subgrid models and those used for \eris{}, in particular their higher
density threshold for star formation and neglect of metal line cooling in gas
with temperature $>10^{4}\, \mathrm{K}$ \citep{Guedes2011}.}

%
%
%

\section{Conclusions} \label{sec:CONCLUSIONS}

We find the following properties of \ins stellar halo stars in three high
resolution SPH simulations of Milky Way analogues:

\begin{enumerate}

\item The \ins stellar halo accounts for 30-40 per cent of the stellar mass
  outside the thin disc \REF{($\mathcal{E}_{\rm E} > 0.8$) and inner spheroid
  (`bulge'; $r>5$~kpc)}. This fraction includes stars that observers may
  classify as belonging to a thick disc.

\item The \ins halo dominates over the \acc stellar halo at $r<30$~kpc in two out of our three simulations. In the third simulation, both \ins and \acc halo stars have almost the same volume density distribution.

\item Between 2 and 30 per cent of the \ins halo comprises stars scattered from near-circular orbits in the plane of the thin disc to more eccentric orbits. These form at lookback times of 5 to 9~Gyr, retain high circular velocities and have relatively narrow MDFs with median $-1 < \mathrm{[Fe/H]} < -0.5$. This component resembles the Milky Way's thick disc, although we note that, in two out of our three simulations, kinematically selected eccentric/thick discs also have a significant contribution from both \acc stars and stars formed from accreted gas.

\item The rest of the \ins halo stars form in gas clouds in the circumgalactic
  halo, on highly eccentric orbits near the disc plane, or in the chaotic
  gas-rich stage of the galaxy's formation ($z>2.5$), before a stable disc is
  established. In one simulation, these halo stars strongly resemble the
  `classical' isotropic, metal-poor, dispersion-supported stellar halo of the
  Milky Way. In the other two simulations, most such halo stars have
  significant rotation in the same sense as the thin disc. 

\item We identify two distinct origins for these stars: gas that has been
  stripped out of satellite galaxies by tides and ram pressure, and gas that is
  incorporated directly into the `smooth' halo of the main galaxy by
  cosmological infall and SN-driven outflow from the central galaxy. 

\item Stars formed \ins from stripped gas have a very similar MDF to the \acc
  stellar halo, because this gas is brought in by the same progenitor
  satellites. Halo stars formed from smoothly accreted gas have a broader MDF
  and are, on average, the most metal poor of the components we identify.

\item The density profile of halo stars formed \ins from
  stripped gas is more concentrated than that of the stars accreted from the
  same progenitor galaxies. This reflects dissipative collapse of
  this stripped gas after it is liberated, rather than differences in how and
  when star and gas particles are stripped, or in the contributions of
  different satellites.

\item In all cases, among the \ins halo stars not scattered from the disc,
  there is almost no difference in the present-day phase-space distribution of
  those formed in stripped and smoothly accreted gas. The correspondence is so
  close that we suggest that, in our simulations, star formation in the hot
  gaseous halo is directly triggered by the passage of dense clumps of
  star-forming stripped gas. 

\item The properties of \ins stars in the solar neighbourhood are
  representative of the \ins halo overall, except that stars formed from
  smoothly accreted gas in this region are notably more metal rich.

\end{enumerate}

Based on these findings, we conclude that essentially \textit{all} halo stars
in our simulation are the result of cosmological accretion and merging, with no
obvious bimodality due to an \ins halo forming through a quasi-monolithic
collapse of enriched diffuse gas. 

\REFB{Comparison with other studies that describe \ins stellar haloes in
hydrodynamical models of Milky Way-like galaxies suggests that a wide variety
of physical mechanisms can be responsible for their formation.  The relative
importance of these processes is particularly sensitive to the implementation
and resolution dependence of subgrid physical prescriptions for star formation
and feedback. An improved understanding of these effects and observational
constraints on star formation in free-floating gas clouds are thus necessary
before definitive conclusions about \ins stellar halo formation can be drawn
from any particular simulation.  This uncertainty is currently the most
important limitation on theoretical predictions for the \ins stellar halo. The
comparison of simulations from different groups will be extremely useful in
this regard, hence our emphasis on simple origin-based definitions for the
different halo components.}

\section*{Acknowledgements}

The authors thank the anonymous referee for their insightful and constructive
suggestions. They are grateful to Takashi Okamoto, who developed our simulation
code, and Patricia Tissera, for useful discussions.  APC is supported by a
COFUND/Durham Junior Research Fellowship under EU grant [267209] and thanks
Liang Gao for support in the early stages of this work under a CAS
International Research Fellowship and NSFC grant [11350110323]. OHP was
supported by NASA grant NNX10AH10G and by NSF grant CMMI1125285. CSF
acknowledges an ERC Advanced Investigator grant COSMIWAY [GA 267291]. This work
was supported by the Science and Technology Facilities Council [ST/L00075X/1]
and used the DiRAC Data Centric system at Durham University, operated by the
Institute for Computational Cosmology on behalf of the STFC DiRAC HPC Facility
(\url{www.dirac.ac.uk}). This equipment was funded by BIS National
E-infrastructure capital grant ST/K00042X/1, STFC capital grant ST/H008519/1,
and STFC DiRAC Operations grant ST/K003267/1 and Durham University. DiRAC is
part of the National E-Infrastructure. 

\bibliographystyle{mn2e}
\bibliography{bibliography}

\appendix

\section{Quantitative comparison with Tissera et al.}
\label{appendix_tissera}

The initial conditions of haloes Aq-C and Aq-D have already been simulated by
another group using different subgrid prescriptions and at slightly lower
resolution, as reported by \citet{Tissera12,Tissera13a,Tissera14}. This
appendix provides a more detailed quantitative comparison of results for these
two haloes to supplement the discussion in Section~\ref{sec:DISCUSSION}. Here
we apply the definitions of components given by \tiss{} to our simulations in
place of the definitions in Section~\ref{sec:SAMPLE}.

Briefly, in section 2.2 of T12, \tiss{} define an \textit{optical radius}
$r_{\mathrm{opt}}$ enclosing 83 per cent of the total galaxy stellar mass. A
\textit{central} stellar component is defined by all stars with binding energy
$E$ less than $E_{\mathrm{cen}}$, the minimum binding energy of stars with
$r>0.5\,r_{\mathrm{opt}}$. Another threshold energy, $E_{\rm inner}$, is
defined as the minimum binding energy of stars with $r>2\,r_{\mathrm{opt}}$;
stars with $E_{\rm cen} < E < E_{\rm inner}$ constitute the \textit{inner halo}
component, and stars with $E > E_{\rm inner}$ the \textit{outer halo}
component\footnote{We find $r_\mathrm{opt}$ values similar to those of T12. The
energy thresholds corresponding to the different components are only slightly
different from those in fig. 1 of T12: we find $(E_{\rm cen}, E_{\rm inner}) =
(-2.48, -1.56) \times 10^{5}$ and $(-1.78, -1.11)\times
10^{5} \,\mathrm{km^{2}s^{-2}}$, for Aq-C and Aq-D respectively.  If we impose
the T12 values of $r_{\mathrm{opt}}$ we find $(E_{\rm cen}, E_{\rm inner}) =
(-2.22, -1.33) \times 10^{5}$ and $(-1.90, -1.25)\times
10^{5} \,\mathrm{km^{2}s^{-2}}$.}. Stars with present-day circularity
$\mathcal{E} > 0.65$ are assigned to the \textit{disc} component, regardless of
their binding energy.  In the inner and outer halo components, stars that form
within the virial radius of the main halo are classified as \textit{disc
heated} if they have $\mathcal{E} > 0.65$ in any output of the simulation.
Since the disc is `hotter' by definition in T12 than in our approach,
relatively fewer disc stars will be classified as `heated'. On the other hand,
the heated disc category in T12 is not restricted to stars scattered from the
$z=0$ disc, which could counter the effect of a lower circularity threshold by
increasing the number of stars assigned to this component\footnote{Section 2.2
of T13 states that their classification, including the identification of the
disc, is only applied at redshifts $z \lesssim 4$. The C15 results reported in
Table~\ref{tab:T_COMPIO} do not include this restriction. If we restrict the
assignment of stars to the heated disc category to those formed at $z \lesssim
4$ (classifying those formed earlier as endo-debris), we find that the only
significant change is to the inner halo heated disc component of Aq-C, reducing
its mass fraction from 13 to 6 per cent.}.

Stars that are not heated disc are classified as \textit{endo debris} if they
form within the virial radius\footnote{In our implementation of these
definitions, membership of the self-bound halo identified by \texttt{SUBFIND} is used as
a proxy for a particle being `within the virial radius'.}  and \textit{debris}
(which we call \textit{accreted}) otherwise.  Importantly, the endo-debris
class includes stars that form bound to satellites within the virial radius
which are later stripped -- in the main text we classify these as accreted
rather than \ins.

\begin{figure}
  \centerline{\includegraphics[width=1.0\linewidth, clip=True, trim=0cm 1cm 0cm 0cm]{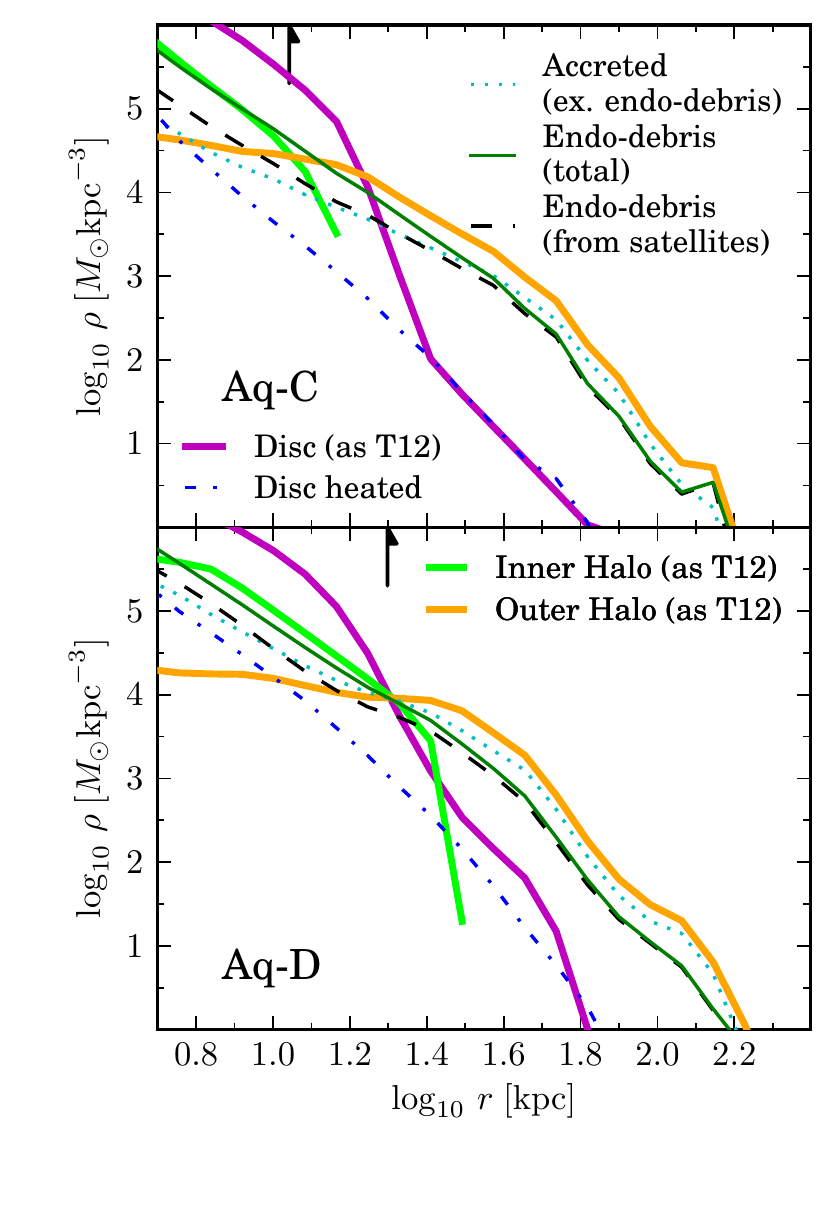}}

\caption{Density profies of stellar components in our simulations Aq-C and Aq-D
according to the definitions of T12 and T13 (see the text). Thin lines show
different categories of halo stars -- accreted (dotted; called debris by T13),
endo-debris (solid) and heated disc (dot--dashed). In addition we show
separately the profile of stars that meet the T13 definition of endo-debris but
which we would classify as accreted, because they form bound to satellites
(long dashed). Thick lines show profiles of all stars in different phase space
`regions' -- disc, inner and outer haloes. Arrows mark the radius $r_{\rm opt}$
defined by T12. The `central' component of T13 is confined to
$\log_{10}{r/\mathrm{kpc}}<0.8$ and hence is not shown.}

\label{fig:t12_density_comparison}
\end{figure}

Table~\ref{tab:T_COMPIO} compares the masses and mass fractions of various
components between the two subgrid implementations and
Fig.~\ref{fig:t12_density_comparison} shows the density profiles of the
different components.  Differences in the properties of the stellar halo need
to be interpreted in the context of large differences in the bulk properties of
the central galaxies (summarized in \citealt{Aquila}). Relative to T12, our
galaxies convert about 50 per cent less of their total baryonic mass to stars
overall\footnote{Our simulations would convert slightly more baryons into stars
at the lower resolution of T12 -- see appendix B.} but still have more massive
discs (by factors of 6.5 for Aq-C and 1.5 for Aq-D). The central (`bulge')
component is less massive in our simulations (by 49 and 28 per cent
respectively). The inner halo component differs even more -- roughly a factor
of 10 less massive in our Aq-C and a factor of 6 less massive in our Aq-D. The
outer halo is most closely comparable, being only slightly less massive in our
simulations.

Given these differences, it is more useful to compare mass fractions than
absolute masses. In the \tiss{} inner halo component, we find a substantially
smaller contribution from scattered disc stars (discussed in the main text) and
a much larger contribution from endo-debris. The balance between \ins and
accreted stars is slightly in favour of the former in our simulations thanks to
this larger endo-debris contribution and a much lower mass of accreted stars in
the same region. As shown in Fig.~\ref{fig:t12_density_comparison}, most of the
endo-debris in our simulation comes from star formation in galaxies after they
have become satellites. In the outer halo, the masses of endo-debris stars in
our simulations are similar to those reported by \tiss{}, but the accreted
stellar mass in this region is roughly two to three times lower, giving a higher
endo-debris fraction. Since endo-debris stars formed in satellites constitute
the majority of endo-debris outside $\sim30$~kpc in our simulations, it is not
surprising that the endo debris and debris (accreted) profiles in
Fig.~\ref{fig:t12_density_comparison} are very similar.

We conclude that the star formation model in our simulations results in
satellites that are more gas rich at accretion and that form stars more
vigorously thereafter, leading, in the language of \tiss{}, to more endo-debris
and less debris. This is consistent with star formation overall being more
efficient in the \tiss{} simulations. 

\REFB{It is interesting in this regard that our simulations and those of
\tiss{} are consistent with the Milky Way satellite luminosity function,
although this comparison involves only a small number of objects.  The gas
content of surviving satellites and dwarf galaxies outside the virial radius
also provides a useful constraint on predictions for the formation of \ins
from accreted gas. In our simulations, we examine the ratio of neutral
($T\lesssim10^{4}\mathrm{K}$) gas mass to stellar mass. We find 3, 8 and 9
satellites at $r<250\kpc$ with $M_{\mathrm{H_{I}}}/M_{\star} > 1\times10^{-4}$ in
haloes C, D and E respectively, only slightly in excess of the same count
around the Milky Way (4) and M31 (5, including M33). Outside this radius, we
find that our simulated dwarf galaxies have a median gas fraction of
$M_{\mathrm{H_{I}}}/M_{\star} \sim 10$, considerably higher than the observed
median for the Local Group, $M_{\mathrm{H_{I}}}/M_{\star} \sim 1$
\citep{McConnachie12}. This supports the idea that, in our models, a low star
formation efficiency before infall may lead to high star formation rates in
stripped gas and in recently accreted satellites.}

As noted in the main text, the high star efficiency of the central galactic
spheroid, presumably fuelled by smoothly accreted gas, might also explain the
substantially lower metallicity of the endo-debris component in T13. Another
point worthy of note is that our simulations predict even lower halo stellar
mass from disc heating than \tiss{}, despite producing more massive discs at
$z=0$.

\begin{table}
  \centering

  \caption{Columns compare results for haloes Aq-C and Aq-D as simulated by
  Tissera et al. (2012, 2013; T12, T13) and in this work (C15), using the
  definitions given \REFB{in} T12 and T13 in all cases.  The first row gives
  the optical radius defined by T12 as enclosing a stellar mass $M_{\star, \rm
  opt}$ (second row) equal to 83 per cent of the total stellar mass of the
  galaxy, $ M_{\star}$.  Rows 3 to 7 give the mass of stars in different
  galactic components of T12, described in the text. Rows 8 to 13 separate the
  inner halo component by origin according to T13, giving the mass of `debris',
  `heated disc' and `endo--debris' stars and their relative fractions. Rows
  14--19 give the same breakdown for the outer halo component.  All quantities
  assume $h=0.73$.}

\label{tab:T_COMPIO}
  
  \begin{tabular}{llrrrr}

                  &           & \multicolumn{2}{l}{Aq-C} & \multicolumn{2}{l}{Aq-D}      \\ 
                  &           & T12  & C15  & T12  & C15   \\
                              \hline 
                              \hline
    $r_{\rm opt}$ & $[\rm kpc]$             & 16.0 & 11.0 & 14.8 & 19.8  \\
    $M_{\rm \star, opt}$ & $[10^{9} \msol]$ & 8.12 & 3.49 & 6.04 & 2.86  \\
    \hline
    \hline
    $M_{\rm disc}$    & $(10^{9} \msol]$          & 2.19 & 14.3 & 12.2 & 18.3  \\
    $M_{\rm central}$ & $(10^{9} \msol]$          & 51.8 & 22.7 & 32.9 & 9.37  \\
    $M_{\rm inner}$   & $(10^{9} \msol]$          & 19.5 & 1.99 & 18.6 & 2.94  \\
    $M_{\rm outer}$   & $(10^{9} \msol]$          & 4.93 & 3.02 & 5.48 & 3.87  \\ 
    $M_{\rm total}$   & $(10^{9} \msol]$          & 78.4 & 42.0 & 69.2 & 34.5  \\

    
    \hline
    \multicolumn{6}{l}{Inner Halo} \\
    \hline

   
    $M_{\rm debris}$ & $(10^{9} \msol]$           & 8.58 & 0.19 & 8.00 & 0.90  \\
    $M_{\rm heated}$ & $(10^{9} \msol]$           & 4.68 & 0.27 & 4.84 & 0.39  \\
    $M_{\rm endo}$   & $(10^{9} \msol]$           & 6.24 & 1.53 & 5.58 & 1.65  \\

    $f_{\rm debris}$ &      \%                      & 44   &  9   & 43 & 31  \\
    $f_{\rm heated}$ &      \%                      & 24   & 13   & 26 & 13  \\
    $f_{\rm endo}$   &      \%                      & 32   & 77   & 30 & 56  \\


    \hline
    \multicolumn{6}{l}{Outer Halo} \\
    \hline
   
    $M_{\rm debris}$ & $(10^{9} \msol)$           & 3.90 & 1.35   & 4.16 & 2.34  \\
    $M_{\rm heated}$ & $(10^{9} \msol)$           & 0.00 & 0.06   & 0.00 & 0.07  \\
    $M_{\rm endo}$   & $(10^{9} \msol)$           & 1.04 & 1.62   & 1.31 & 1.47  \\

    $f_{\rm debris}$ &    \%                        & 79 & 45  & 76 & 60  \\
    $f_{\rm heated}$ &    \%                        & 0  &  2  &  0 &  2  \\
    $f_{\rm endo}$   &    \%                        & 21 & 53  & 24 & 38  \\
       
    \hline
  \end{tabular}
\end{table} 

\section{Numerical resolution}
\label{appendix_restest}

We have reported three mechanisms responsible for the generation of \ins
halo stars in our simulations -- disc heating, star formation in stripped gas
and star formation in smoothly accreted gas --  and discussed similarities and
differences with other simulations at comparable resolution.  All three
mechanisms are all likely to be sensitive to numerical resolution, for
different reasons. 

The disc heating dynamics in $N$-body simulations and their resolution
dependence have been studied extensively
\citep[e.g.][]{Kazantzidis:2008aa,Read:2008aa,Purcell2010}; in ab initio
simulations there is an additional dependence on the subgrid physics governing
the formation and structure of the disc \citep[e.g][]{DeBuhr:2012aa,Aumer13a}.
This dependence is now being tackled directly, with some success, by recent
generations of hydrodynamical simulations \citep{Aquila, Aumer13b,
Schaye:2015aa, Snyder:2015aa}. The behaviour of subgrid star formation models
is less well understood in the regime of dwarf galaxies, where there are fewer
constraints -- the most important being comparison to the satellite luminosity
functions of the Milky Way and M31 \citep[e.g.][]{Parry2012, Zolotov:2012aa,
Sawala:2013aa}.  The star formation histories of satellites determine their
contribution to the accreted halo and their gas content at accretion, which
provides the reservoir for star formation in tidal and ram-pressure streams.
The formation of stars in gaseous streams and clumps depends on the treatment
of hydrodynamic interactions within circumgalactic haloes of hot gas
\citep[e.g.][]{Maller:2004aa}. Moreover, although these hot gas haloes are a
robust prediction of the CDM model, their detailed properties in simulations
are known to be sensitive to the treatment of thermal instabilities and
galactic winds (and, potentially, active galactic nuclei).  Subgrid star
formation models are typically calibrated in the context of a dense cold
ISM, with star formation in unstable pockets of halo gas most
often considered a `nuisance' effect \citep[e.g.][]{Kaufmann06,Keres12,
Parry2012,Hobbs13}.

Since our main purpose is to present the phenomenology of particular sets of
hydrodynamical simulations, rather than make predictions for specific
observations, we have not carried out detailed convergence studies to address
these issues. It is possible that the numerical effects given above are
dominant over physical behaviour in all cases; hence, tests of convergence will
be absolutely essential for any robust predictive model of \ins stellar halo
formation.  

In this respect we believe the formation of stars from `smoothly accreted' hot
gas to be the least robust of the mechanisms we identify, having three likely
sources of resolution dependence: the behaviour of the subgrid star formation
model, the treatment of thermal instabilities and the classification of
particles as `smoothly' accreted. It is not at all clear a priori which of
these dependences is most important for convergence. 

\begin{figure}
  \centerline{\includegraphics[width=1.0\linewidth, clip=True, trim=0cm 0.5cm 0cm 0cm]{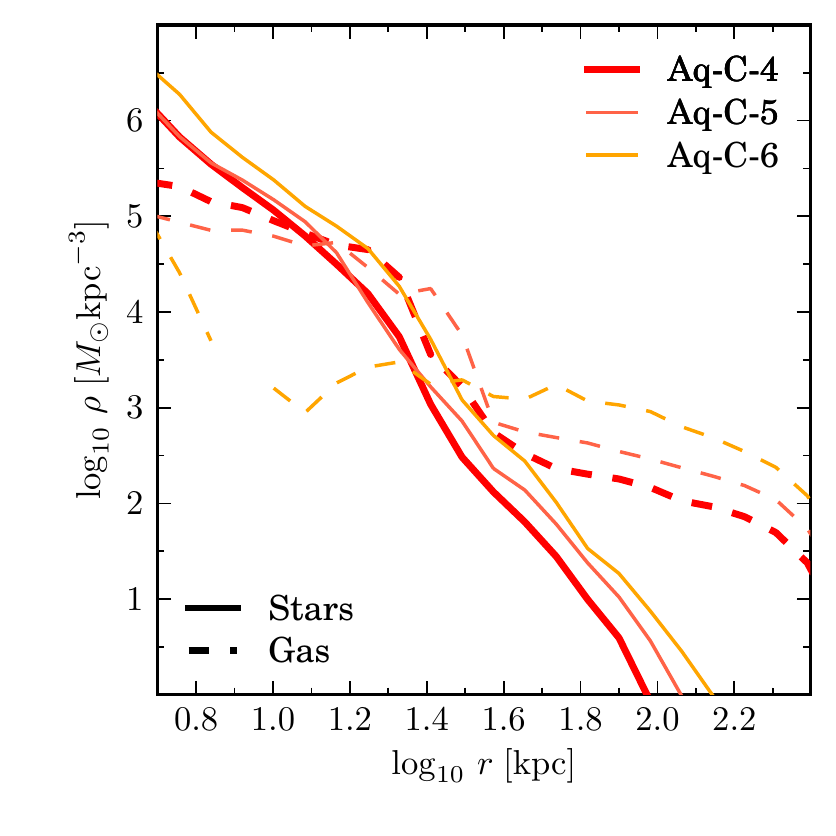}}

\caption{Present-day mass density profiles of smoothly-accreted gas particles
(dashed) and all star particles formed from smoothly accreted gas (solid), in
halo Aq-C at our default resolution level (L4). Thinner lines correspond to
lower resolution levels L5 and L6 as described in the text.}

\label{fig:res_test}
\end{figure}

In Fig.~\ref{fig:res_test} we illustrate the resolution dependence of the density profile
of stars formed from gas accreted smoothly by the main halo. The solid red line
corresponds to Aq-C at our default resolution level (level 4), including the
disc and the stellar halo.  Solid lines of lighter colours correspond to
simulations from the same initial conditions with progressively lower
resolution (levels 5, particle mass $m_{\rm p} = 2.11\times10^{6}$, and 6, $m_{\rm p} =
1.69\times10^{7}$; these simulations are also discussed in
\citealt{Parry2012}).  

As resolution increases (from L6 to L4), the mass of stars formed from smoothly
accreted gas is reduced at all radii. Although the profile of disc stars associated with this component appears
to converge at resolution L5, this is not the case for the stellar halo.
In a region $r>30\kpc$, we find $4\times10^{8}\msol$ of stars formed
from smoothly accreted gas at L6, representing 11 per cent of the stellar mass
and 0.4 per cent of all baryons in the same region. With increasing resolution,
we see a reduction in the total mass of this component by $\sim50$ per cent
for each level, a similar change in its mass relative to all stars (from 11 to
4 per cent, a 43 per cent reduction in this fraction per level) and a somewhat
smaller reduction in its mass relative to all baryons (17 per cent in this
fraction per level). Dashed lines in  Fig.~\ref{fig:res_test} tell a similar
story for the gas that is smoothly accreted but {\rm not} converted to stars --
this is always the dominant baryonic mass component at $r>30\kpc$.

The absolute change in the mass of halo stars formed from smoothly accreted gas
is explained only in part by a reduction in the total baryonic mass of the halo
by $\sim30$ per cent from L6 to L4 ($\sim36$ per cent for $r>30\kpc$; this can
be compared to a 6 per cent decrease in total mass bound to the main
substructure). Another significant factor must therefore be that the
classification of a gas particle as `smoothly accreted' may also not be
converged, because gas associated with DM substructures resolved only
at higher resolution will be classified as `smoothly' accreted at lower
resolution. This is illustrated by the results of \citet{Wang2011}, who examine
the total mass (baryons and DM) accreted smoothly on to the main halo
in collisionless versions of the same initial conditions we simulate. Their
fig. 9 shows a decrease of $\sim13$--$15$ per cent in the faction of
`smooth' accretion from L5 to L4 (depending on how `smooth' accretion is
defined).  If all DM subhaloes supplied baryons to the main halo in
the universal ratio, we would expect a similar reduction in the fraction of gas
classified as smoothly accreted.

We conclude that, at lower resolution, more baryons are retained by the main
halo and relatively more of these baryons are considered as smoothly accreted.
The overall efficiency with which smoothly accreted gas is converted to stars
also varies systematically with resolution, although (in our simulations) this
is a minor contribution to the change of \ins stellar halo
mass, relative to the first two effects. 

\label{lastpage} \end{document}